\documentclass[fleqn,usenatbib]{mnras}

\usepackage[utf8]{inputenc}
\usepackage[T1]{fontenc}
\usepackage{newtxtext,newtxmath}
\usepackage{algorithmic}
\usepackage[lined,ruled,linesnumbered]{algorithm2e}
\usepackage{graphicx}	
\usepackage{amssymb}	
\usepackage{flushend}

\newcommand{\bb}{\boldsymbol{b}}
\newcommand{\bD}{\boldsymbol{D}}
\newcommand{\bd}{\boldsymbol{d}}

\newcommand{\bod}{\overline{\boldsymbol{d}}}
\newcommand{\be}{\boldsymbol{e}}

\newcommand{\bH}{\boldsymbol{H}}

\newcommand{\bI}{\boldsymbol{I}}

\newcommand{\bm}{\boldsymbol{m}}

\newcommand{\bnabla}{\boldsymbol{\nabla}}

\newcommand{\bp}{\boldsymbol{p}}

\newcommand{\bs}{\boldsymbol{s}}
\newcommand{\bt}{\boldsymbol{t}}
\newcommand{\bT}{\boldsymbol{T}}
\newcommand{\bU}{\boldsymbol{U}}
\newcommand{\bu}{\boldsymbol{u}}

\newcommand{\bV}{\boldsymbol{V}}
\newcommand{\bv}{\boldsymbol{v}}
\newcommand{\bW}{\boldsymbol{W}}
\newcommand{\bw}{\boldsymbol{w}}
\newcommand{\bx}{\boldsymbol{x}}
\newcommand{\bovx}{\overline{\boldsymbol{x}}}
\newcommand{\bX}{\boldsymbol{X}}
\newcommand{\btX}{\boldsymbol{\tilde{X}}}
\newcommand{\bY}{\boldsymbol{Y}}
\newcommand{\by}{\boldsymbol{y}}
\newcommand{\bovy}{\overline{\boldsymbol{y}}}

\newcommand{\mybtheta}{\boldsymbol{\theta}}

\title[MUFFIN deconvolution algorithm]{A parallel \& automatically tuned algorithm for multispectral image deconvolution}

\author[R. Ammanouil et al.]{ R. Ammanouil$^{1,2}$, A. Ferrari $^2$, D. Mary$^2$, C. Ferrari$^2$ and F. Loi$^3$ 
\\
$^1$ College of Engineering and Technology, American University of the Middle East, Kuwait \\
$^2$ Université Côte d'Azur, Observatoire de la Côte d'Azur, CNRS, Lab. J.-L. Lagrange, France  \\
$3$ Dipartimento di Fisica e Astronomia, Università degli Studi di Bologna, Italy
}

\date{Accepted XXX. Received YYY; in original form ZZZ}

\pubyear{2019}

\begin{document}
\label{firstpage}
\pagerange{\pageref{firstpage}--\pageref{lastpage}}
\maketitle

\begin{abstract} 
In the era of big data, radio astronomical image reconstruction algorithms are challenged to estimate clean images given limited computing resources and time. This article is driven by the need for large scale image reconstruction for the future Square Kilometre Array (SKA), which will become in the next decades the largest low and intermediate frequency radio telescope in the world. This work proposes a scalable wideband deconvolution algorithm called MUFFIN, which stands for ``MUlti Frequency image reconstruction For radio INterferometry''. MUFFIN estimates the sky images in various frequency bands given the corresponding dirty images and point spread functions. The reconstruction is achieved by minimizing a data fidelity term and joint spatial and spectral sparse analysis regularization terms. It is consequently non-parametric w.r.t. the spectral behaviour of radio sources. MUFFIN algorithm is endowed with a parallel implementation and an automatic tuning of the regularization parameters, making it scalable and well suited for big data applications such as SKA. Comparisons between MUFFIN and the state-of-the-art wideband reconstruction algorithm are provided. 
\end{abstract}

\begin{keywords}
techniques: image processing - techniques: interferometric  - methods: data analysis - methods: numerical  
\end{keywords}

\section{Introduction}

The new generation of radio interferometers is characterized by very wide fields of view,  high resolution, huge sensitivity and very large bandwidth. The future Square Kilometre Array (SKA) is emblematic of these extreme instruments. In its final phase of construction, SKA will consist of up to one million dipole antennas and two thousands parabolic antennas \citep{dewdney2009square,dewdney2013ska1}. These antennas,  comprising a collecting area of approximately one square kilometer, will be located in Australia and South Africa. The SKA radio telescope is expected to provide images of the sky in the giga pixel range with \mbox{(sub-)}arcsec angular resolution, fields of view up to 200 $\text{deg}^2$ and more sensitivity than any other present-day instrument with a dynamic range covering up to seven orders of magnitude (see e.g. Table 1 in \citet{dewdney2013ska1}). 

The work presented here addresses the large-scale and high-precision radio-interferometric image reconstruction problem. Radio-interferometric data, the visibilities, are the cross-correlations between voltages from pairs of antennas. In the ideal case, two receivers with baseline $\bb$ observing in a narrow frequency band $c/\lambda$ measure a complex visibility equal to the Fourier spectrum of the intensity distribution of interest at spatial frequency $\bb/\lambda$. The partial measurement of the Fourier plane is what makes the image reconstruction strongly ill posed. In practice, the  measurement model corresponds to a linear map from the image domain to the visibility space, but includes multiple distortions such as: a)  direction-dependent effects (DDE) related to wide-field effects, non-coplanarity of the baselines, the ionospheric perturbations and associated Faraday rotation; b) antenna level effects such as the primary beam patterns, element-based gains and c) computational approximation of operators such as the degridding required to use Fast Fourier Transform algorithms, see for example \citep{wijnholds2010challenges,rau2009,bhatnagar2008}. Compensations of the perturbations a) and b) rely on alternating a (self-)calibration step and an image reconstruction  step. This paper concentrates on the image reconstruction step. 

The imaging task is realized by solving either directly the inverse problem related to the calibrated visibility measurements, or the inverse problem in the image domain after applying the adjoint measurement model to the visibilities. The latter approach relies on the construction of a `dirty image' from calibrated visibilities. The dirty image is related to the sky image by a convolution if the DDEs are properly compensated. This is the approach followed in the classical major/minor loop  \citep{jongerius2014,tasse2013applying} and in the facet frameworks \citep{kogan2009,vanWeeren2016}; see \citep{tasse2018} for a detailed comparison between these two techniques. The main advantage of the image domain processing is not only the computing efficiency related to the fact that the direct and adjoint operators are simple convolutions, but also the amount of data to process: the dirty image size equals the reconstructed sky image size while the number of raw visibilities can be several order of magnitudes the number of pixels in the sky image.     

Interestingly, sparsity was early recognized as a powerful tool for radio-interferometric image reconstruction and has lead to the most populated family of imaging algorithms. Their patriarch is the CLEAN algorithm \citep{hogbom1974aperture}, which expresses and exploits the sparsity of the sky intensity distribution in the canonical basis. Variants of CLEAN as \citet{Cornwell2008} generalize this approach to extended objects. More recently, various algorithms based on the theory of compressed sensing \citep{donoho2006compressed,candes2008introduction} and convex optimization emerged for the reconstruction of radio interferometric images from the calibrated visibilities. These algorithms exploit the fact that the image has a sparse representation in some appropriate transform domain through the use of sparsifying regularizations; see for example  \citep{wiaux2009compressed, mcewen2011compressed,vijay2017fourier}. A greedy alternative to the minimization of a regularized cost function was proposed by \citet{dabbech15}. Sparse image representations using  wavelet dictionaries has indisputably shown substantial improvement with respect to  classical CLEAN-based approaches, as demonstrated  by \citet{garsden15} using LOFAR data and \citet{dabbech2018} using VLA data. The price to pay is an increased computational cost. As a result, a lot of effort has gone into designing scalable methods able to cope with large data sets. E.g., the authors of \citep{Onose16,onose2016randomised} achieve parallel computations by splitting the visibilities into multiple blocks and exploiting randomization over data blocks at each iteration.

The SKA will observe the sky in wide sub-bands (covering all together from $50$ MHz to $14$ GHz) and the reconstruction of both spatial and spectral behavior of radio sources is essential  to characterize the astrophysical origin of the detected  radiation, see e.g. \citet{krausb}. As a result, multi frequency  reconstruction algorithms are required in order to fully exploit the spectral and spatial resolution of the SKA telescope. Taking the path of CLEAN, the Multi-Frequency Synthesis (MFS) \citep{Conway90}  and the Multi-Scale Multi-Frequency Synthesis (MS-MFS) \citep{rau2011multi}   deconvolution algorithms  were the first wide-band (WB) radio-interferometric  reconstruction algorithms. At the time of writing, still relatively few such WB reconstruction algorithms  exist  \citep{bajkova2011multifrequency,Wenger2014,Junklewitz2015,abdulaziz2016low}. 

In \citet{rau2011multi,bajkova2011multifrequency,Junklewitz2015} the spectrum of each pixel is modeled as a (generalized) power law, or its Taylor expansion. These `semi-parametric' methods rely on spectral models and thus clearly offer many advantages such as estimation accuracy when the model is appropriate. However, across the broad frequency coverage of current radio facilities, radio sources exhibiting complex spectral shapes (not simple power laws) are expected. For instance,  \citet{Kellerman74} evidences that some sources may exhibit one or more relative minima, breaks and turnovers. For the new generation of low frequency telescopes  such as LOFAR, recent studies have also shown that second order broadband spectral models are often insufficient \citep{Scaife12}. In addition to that, not assuming any spectral shape for radio sources in the deconvolution step can be useful for mutli-wavelength polarization studies, as well as for simultaneously identifying spectral lines within wide-band continuum observations. However, few WB reconstruction algorithms consider the general case of reconstructing the spatio-spectral `image cube' without constraining the spectral variation to a power law. In particular, few WB reconstruction algorithms have opted instead for a sparse regularization of the spectra. Among these, \citet{Wenger2014} formulate the problem as an inverse problem with a smooth spectral regularization allowing for local deviations; \citet{abdulaziz2016low} constrain the sky image cube to be low rank in order to enforce that only few sources, each with a distinct spectral signature, can exist in the field of view. This approach  requires a singular value decomposition of the whole sky image cube at each iteration, so computational costs can become an issue. Finally, WB image reconstruction has also recently been tackled as a joint blind source separation and reconstruction problem by \citet{jiang2017joint}.

The present article focuses on total intensity wide-band and HI imaging task within the minor loop. Section 2 derives the proposed algorithm  which is nonparametric in the sense that it does not rely on a particular assumption of the spectral structure, but rather on a sparse representation of the spectra in a dictionary. The spectral dimension critically blows up the size of the inverse problem and a particular  effort is dedicated to the scalability of the algorithm. This results in an algorithm where most computationally intensive steps, such as the spatial wavelets decompositions and the associated adjoint operators, can be computed in parallel for each wavelength. Given the execution time for a large scale reconstruction, the selection of the regularization parameters is particularly critical. Section 3 proposes an automatic tuning strategy based on the Stein Unbiased Risk Estimator. Finally, section 4 presents simulation results which demonstrate the effectiveness of the algorithm. 

It is worthy to note that, whereas simulations in section 4 are realized in the deconvolution framework, the proposed algorithm can be used in the same way to reconstruct the sky image cube from dirty images or visibilities, depending on the direct operator. 

\section{Inverse problem}
In this section, we first formulate the multi-frequency image deconvolution task as a regularized and constrained least squares optimization problem. We then propose and discuss a primal-dual algorithm for solving the aforementioned optimization problem.

\subsection{Optimization problem}
Let $\bx_{\ell}{\bf{^\star}}$ and $\by_{\ell}$ be $N\times 1$ column vectors collecting the unknown (and sought-after) sky intensity image and the (measured) dirty intensity image respectively. The subscript $\ell$ in the notations $\bx_\ell{\bf{^\star}}$, $\by_\ell$ and in all following matrix and vector notations refers to the wavelength band $\lambda_{\ell}$, with $\ell=1,\ldots,L$ and $L$  the  number of spectral channels. At each wavelength band, the dirty image $\by_{\ell}$ is related to the sky image $\bx_{\ell}{\bf{^\star}}$ by a linear mapping and contaminated with noise. This takes the following form:
\begin{equation}
\label{model}
\by_{\ell} = \bH_{\ell} \bx_{\ell}^\star  + \bw_{\ell}, \quad \ell=1,\ldots, L
\end{equation}
where $\bH_\ell$ represents a convolution by the point spread function (PSF), and $\bw_{\ell}$ is a perturbation vector accounting for noise and modeling error. The perturbation vector $\bw_{\ell}$ is assumed to be Gaussian, independent and identically distributed (i.i.d) with zero mean. Without loss of generality we will assume that the variance of the noise does not depend on the wavelength and is equal to $\sigma^2_w$.  More sophisticated correlation structures can be taken into account provided that the covariance matrix is known. Given the dirty images and the PSF, Eq.~(\ref{model}) defines an ill-posed deconvolution problem. The problem is ill-posed owing to the partial coverage of the Fourier plane, i.e. due to the fact that $\bH_{\ell}$ is rank deficient.  A meaningful solution can be obtained in a cost minimization framework by incorporating regularizations and constraints reflecting our prior knowledge on the unknown variables $\bx_1{\bf{^\star}},\ldots,\bx_{L}{\bf{^\star}}$, the true sky intensity images.  Let  $\bx_1,\ldots,\bx_{L}$ denote the optimization variables corresponding to $\bx_1{\bf{^\star}},\ldots,\bx_{L}{\bf{^\star}}$, and let $\bX$ denote the $N\times L$ matrix $ \bX := [\bx_1,\ldots,\bx_{L}]$ obtained by stacking the image vectors next to each other. Let $\bx^1, \ldots, \bx^N$ denote the rows of $\bX$ which represent the pixels spectra. Hence, $\bX$ can be alternatively obtained by stacking the pixels spectra on top of each other, i.e. $\bX:=[{\bx^1}^\top, \ldots, {\bx^N}^\top]^\top$. With these notations, we propose to minimize the following cost function: 
\begin{equation} 
\label{probopt}
\text{argmin}_{\bX}\; \sum_{\ell=1}^{L} \frac{1}{2 }  \|\by_{l} - \bH_{\ell} \bx_{\ell} \|^2 + \boldsymbol{1}_{\mathbb{R}^+}(\bX) + h(\bX),
\end{equation}
with 
\begin{equation}
h(\bX) := 
\sum_{\ell=1}^{L}  \mu_\ell \| \bW_s \bx_{\ell} \|_{1} + \sum_{n=1}^{N} \gamma_n \| \bW_\lambda \bx^{n} \|_1.	 
\label{RegAll}
\end{equation}
The first term in the cost function in equation \eqref{probopt} is a data fidelity term penalizing the discrepancy between the measurements and the estimated model. The second term in equation \eqref{probopt} is a positivity constraint enforcing each element in $\bX$ to be positive, more precisely $\boldsymbol{1}_{\mathbb{R}^+}(\bX)=0$ if $\bX \geqslant 0$ (where the inequality  is element-wise) and $\boldsymbol{1}_{\mathbb{R}^+}(\bX)=\infty$ otherwise. The third term in equation \eqref{probopt}, developed in equation \eqref{RegAll},  is a spatial-spectral regularization. More precisely, $h(\bX)$ is the sum of a spatial regularization imposed on the images at all the wavelength bands and a spectral regularization imposed on the spectra at all the pixels. The matrices $\bW_s$ and $\bW_\lambda$ represent the spatial and spectral decomposition operators respectively. The scalars $\mu_\ell \in \mathbb{R}^{+}$ for $\ell=1, \ldots L$ and $\gamma_n \in \mathbb{R}^+$ for $n=1, \ldots N$ are the spatial and the spectral regularization weights respectively. 

The spatial and spectral regularizations in equation \eqref{RegAll} are sparse analysis regularizations \citep{elad2007analysis}. The sparse analysis approach consists in penalizing the $\ell_1$-norm of the correlations between the atoms in a dictionary, i.e. $\bW_s$ and $\bW_\lambda$ in the case of \eqref{RegAll}, and the images, i.e. $\bx_\ell$ for $\ell=1, \ldots, L$ and the spectra, i.e. $\bx^{n}$ for $n=1, \ldots, N$. Penalizing the $\ell_1$-norm of these decompositions promotes a sparse correlation between the atoms in the dictionary and the variables. Hence, it encourages finding a solution that is not correlated to some atoms of the dictionary and conversely more correlated to others. The dictionaries can be orthonormal transforms or more generally redundant transforms also referred to as over-complete dictionaries. Possible choices for  $\bW_s$  are (concatenations of) Daubechies wavelets \citep{Purify,Onose16} or the isotropic undecimated wavelet transform (IUWT) \citep{garsden15,starck2010astronomical,dabbech15} that have proved their efficiency for various image processing tasks in astronomy. One possible choice for  $\bW_\lambda$  that will be illustrated in the experiments is the Discrete Cosine Transform (DCT), a Fourier related transform. The DCT spectral regularization assumes that the DCT is compact, an assumption widely used in the sparse reconstruction literature. 

Other types of regularizations based on sparse representations, such as synthesis priors or more generally representations based on 3D spatio-spectral dictionaries, can be of course considered. See  \citep{Lanusse2014} for a general discussion on 3D wavelets and e.g. \citep{Paris2013} where a spatio-spectral dictionary is derived for a specific detection task. The drawback of these regularizations is computational: contrarily to (\ref{RegAll}), they couple space and frequency, i.e. the lines and columns of $\bX$, and prevent an efficient implementation of (\ref{probopt}), e.g. with parallel steps. This point will be clarified in the next section. Finally, it is essential to note that if $\forall \ell$, $\gamma_\ell=0$, the minimization problem in Eq. (\ref{probopt}) becomes separable w.r.t. $\bx_\ell$ and reduces to $L$ deconvolution problems that can be solved independently at each wavelength $\lambda_\ell$.

The regularization parameters $\mu_\ell$ for $\ell=1, \ldots, L$ and $\gamma_n$ control the strength of the regularization term w.r.t. the data fidelity term. A large value of $\mu_\ell$ tends to over-smooth structures at wavelength $\ell$, while a small value of $\mu_\ell$ leads to under-regularization. Similarly, a large value of $\gamma_n$ tends to over-smooth spectra at pixel $n$. See, e.g. \citep{Renard2011}  in the case of a single regularization parameter for monochromatic optical interferometry. Although the use of a different regularization parameter per pixel and  band could improve the estimation performance compared to the use of one regularization parameter per regularization type, the tuning of $(L+N)$ regularization parameters is  in practice very challenging. For this reason, we propose to use two regularization parameters and assign a different (fixed) weight per pixel and per band. By doing this, we reduce the number of regularization parameters from $(L+N)$ to $2$ and we take into account the spatial and spectral variability of the images and spectra respectively. We adopt the following model for the regularization parameters: 
\begin{equation}
\left\{ \begin{array}{ll}
  \gamma_n & = \gamma \alpha_n  , \quad n=1,\ldots, N\\
  \mu_\ell & = \mu \beta_\ell  , \quad \ell=1,\ldots, L \\ 
\end{array}\right. \label{2regpar}
\end{equation}
with 
\begin{equation}
\left\{ \begin{array}{ll}
  \alpha_n & = \frac{1}{\| \bW_\lambda (\bx^n)_0\|_2 + \varepsilon} \\
  \beta_\ell & = \frac{1}{\| \bW_s (\bx_\ell)_0 \|_2 + \varepsilon} \\ 
\end{array}\right.
\end{equation}
 where $(\bx^n)_0$ and $(\bx_\ell)_0$ correspond to the spectra and images of the sky image used for initializing the algorithm. The intuition for the weights setting is similar to the one in re-weighted $\ell_1$ minimization schemes \citep{Cands2008} where the strength of the regularization parameter is inversely proportional to the norm of the corresponding regularized norm. {The intuition for this setting is that the value of the regularization parameter should be high whenever the decomposition coefficient is close to zero in order to encourage sparsity, and it should be small otherwise.} Nevertheless, in re-weighted $\ell_1$ minimization schemes, the weights may be updated at each iteration or after a certain number of iterations using the current estimates for $\bx^n$ and $\bx_\ell$. In this work, and in the simulations, we consider fixed weights computed from a coarse initial estimation of $\bX$.   
 
 Finally, note that the proposed optimization problem in equation \eqref{probopt} is convex but does not admit a closed form solution.  In order to find the minimizer of \eqref{probopt}, we need to resort to an iterative procedure as described in the next section.

\subsection{Optimization algorithm}
\label{sec:optimalg}

The proposed optimization problem described in equations \eqref{probopt} and \eqref{RegAll} is the sum of a smooth and differentiable function (the LS term),  and three non-smooth and non-differentiable functions (the indicator function, the sparse analysis spectral regularization, and the sparse analysis spatial regularization). Nevertheless, the non-differentiable functions, the indicator function and the two sparse analysis regularizations, are simple functions in the sense that their proximity operators \citep{combettes2011} have closed form representations. In order to solve the proposed optimization problem, we resort to the primal dual splitting algorithm proposed by \citet{Vu11,Condat} also known as the V{\~u}-Condat algorithm. The V{\~u}-Condat algorithm is well suited for optimization problems similar to the one proposed in this work i.e. problems that consist of a sum between a smooth function and several proximable functions involving linear operators. Note that other splitting algorithms could have been used such as the Alternating direction Method of Multipliers ADMM \citep{boyd2011distributed}, Simultaneous Direction of Multipliers SDMM \citep{setzer2010deblurring,combettes2011} or the primal dual splitting algorithm proposed by \citet{combettes2012primal}. These algorithms operate by splitting the objective function to be minimized into smaller and simpler functions that are dealt with individually. However, the V{\~u}-Condat algorithm  is more appealing compared to ADMM and SDMM because it does not involve any implicit inversion of the operators $\bI+\alpha\bW_\lambda^\top\bW_\lambda$ and $\bI+\beta\bW_s^\top\bW_s$ which can be very large and cumbersome to invert, see for example the work by \citet{meillier2018distribution}. It can be considered as a generalization of Chambolle-Pock algorithm \citep{Chambolle2010} which includes a differentiable term and multiple analysis regularizations.

V{\~u}-Condat algorithm is not symmetric with respect to the primal and dual variables. Switching the roles of these variables gives rise to two different versions. A straightforward application of the first V{\~u}-Condat algorithm \citep[Algorithm 3.1]{Condat}  to the optimization problem described in equations  \eqref{probopt} and \eqref{RegAll}  with no relaxation ($\rho_n=1$) leads to algorithm \ref{MUFFIN}. These choices are discussed at the end of this subsection. We refer to algorithm \ref{MUFFIN} as MUFFIN for MUlti Frequency image reconstruction For radio INterferometry. Algorithm \ref{MUFFIN} requires using the proximity operator of the positivity indicator function, namely the projection on the positive orthant defined as: 
\begin{equation}
(u)_+ :=
\left\{ \begin{array}{ll}
  u & \quad \text{if} \quad u > 0\\
  0 & \quad \text{otherwise.}\\ 
\end{array}\right.
\label{projpos}
\end{equation}
The proximity operator of the $\ell_1$-norm is the well-known soft thresholding operator. Algorithm \ref{MUFFIN} requires using the proximity operator of the Fenchel conjugate of the $\ell_1$-norm function which can be easily computed from the soft thresholding operator using Moreau's identity. This leads to the saturation function defined as:
\begin{equation}
\text{sat}(u) :=
\left\{ \begin{array}{ll}
  -1 & \quad \text{if} \quad u < -1\\
  1 & \quad \text{if} \quad u > 1\\ 
  u &  \quad \text{if} \quad |u| \leq 1.
\end{array}\right.
\label{sat}
\end{equation}
Note that both $(\cdot)_+$ and $\text{sat}(\cdot)$ used in algorithm \ref{MUFFIN} are applied element wise when the input is a matrix. The matrices $\bD_\bmu$ and $\bD_\bgamma$ used in algorithm \ref{MUFFIN} are diagonal matrices such that $\bD_\bmu := \text{Diag}(\mu_1, \ldots, \mu_L)$, and $\bD_\bgamma := \text{Diag}(\gamma_1, \ldots, \gamma_N)$, and the parameters $\sigma>0$ and $\tau>0$ are parameters of the algorithm to be fixed appropriately in order to guarantee convergence. More precisely, they must be fixed such that: 
\begin{equation}
\tau \left( \frac{\beta}{2} + \sigma \left(\sum_{\ell=1}^L\mu_\ell^2 \|\bW_s\|^2+\sum_{n=1}^N \gamma_n^2\| \bW_\lambda\|^2\right) \right)<1,
\end{equation}
where $\beta$ is the Lipschitz constant of the gradient of the LS term in equation \eqref{probopt}.

Hereafter, we briefly describe each step in Algorithm~\ref{MUFFIN}. In equation \eqref{grad1}, the gradient of the LS term is computed. This consists of applying the linear operator $\bH_\ell^\dagger$ on the residue of the model for $\ell = 1, \ldots, L$. In equation \eqref{xt1}, $\btX$ is computed. The adjoint operators $\bW_s^{\dagger}$ and $\bW_\lambda^{\dagger}$ are applied on the dual variables $\bU$ and $\bV^\top$. More precisely, the third term is where the operator $\bW_s^{\dagger}$ is applied on $\bU$, and the fourth term is where the operator $\bW_\lambda^{\dagger}$ is applied on $\bV^\top$ ($\bV\bW_\lambda=(\bW_\lambda^{\dagger}\bV^\top)^\top$). The primal variable $\bX$ is updated using the gradient and the aforementioned inverse transforms, and a projection on the positive orthant is then applied. The result is then assigned to the intermediate primal variable $\btX$. In equation  \eqref{updateu1}, the first dual variable $\bU$ is computed. The linear operator $\bW_s$ is applied on $(2\btX-\bX)$. The dual variable $\bU$ is updated using previously computed transform, and the saturation function is then applied. The result is assigned to the dual variable $\bU$. In equation \eqref{updatev1}, similarly to the previous step, the second dual variable $\bV$ is computed. The linear operator $\bW_\lambda$  is applied on  $(2\btX-\bX)^\top$. The dual variable $\bV$ is updated using previously computed transform,, and the saturation function is then applied. The result is assigned to the dual variable $\bV$. In equation \eqref{updatex1}, $\bX$ is updated by simply assigning it to $\btX$. Finally, note how the proposed iterative algorithm~\ref{MUFFIN} activates separately: the gradient of the LS term, i.e. the linear operator $\bH_\ell^\dagger$ for $\ell = 1, \ldots, L$, the linear operators $\bW_s$, $\bW_\lambda$, and their adjoint operators $\bW_s^{\dagger}$,  and $\bW_\lambda^{\dagger}$. For example, the spatial and spectral operators are applied on images and spectra respectively in separate steps, see for example $\bW_s\bX$ in eq. \eqref{updateu1} and $\bW_\lambda\bX^\top$ in eq. \eqref{updatev1}. Finally, for a better understanding of the derivation of algorithm~\ref{MUFFIN}, the reader can refer to Algorithm 3 (section 4) in \citep{Condat}.  

From  \eqref{updateu1}, the size of $\bU$ is the size of $\bW_s\bX$ and consequently several times the size of the cube of images $\bX$ when $\bW_s$ is a redundant transform such as a union of orthogonal wavelets or an IUWT decomposition. Symmetrically to \eqref{updateu1}, the second version of the V{\~u}-Condat algorithm \citep[Algorithm 3.2]{Condat} requires, for the update of $\bX$, the update of $\bU$ in the current iteration and in the previous iteration. Keeping these two variables in memory can be too cumbersome when $\bX$ is large and when the spatial decomposition is redundant. For this reason, we privileged the first version of the algorithm \citep[Algorithm 3.1]{Condat}. Similarly, in order to reduce memory requirements, the relaxed version of the algorithm is not used ($\rho_n=1$).

\subsection{Parallel implementation} \label{sec:parallel}

The parallel implementation consists in parallelizing the steps that are carried on the images at different wavelength bands separately. The advantage of this parallelization is twofold. First the memory load can be distributed among several slave nodes, a slave node only stores the images at certain wavelength bands and not at all wavelength bands. Second, most of the computations and most interestingly the most expensive ones (such as the application of $\bW_s$ and $\bW_s^\dagger$) can be also carried out in parallel. More precisely, a group of $L$ slave nodes, where each slave node is assigned a band, handles the gradient computation, the application of the spatial operator $\bW_s$ and its adjoint operator $\bW_s^\dagger$. 

The parallel implementation of algorithm \ref{MUFFIN} is described in algorithm \ref{MUFFINPar}. The computations in equations \eqref{grad1}-\eqref{updateu1}, and \eqref{updatex1}, except for the last term in equation \eqref{xt1} are computed in a parallel manner, for each band. The parallel computations of equations \eqref{grad1}-\eqref{updateu1}, and \eqref{updatex1} in algorithm \eqref{MUFFIN} are described in equations \eqref{updatenode2}-\eqref{stpu} in algorithm \ref{MUFFINPar}.  After the slave nodes perform there computations in parallel, each node sends the new value of $\bx_\ell$ and $\tilde{\bx}_\ell$ to the master node.  The master node centralizes the computation of the last term in equation \eqref{xt1} and equation \eqref{updatev1}, since these computations operate on spectra rather than images. It is worthy to note that if $\forall n$ $\gamma_n=0$, i.e. the spectral regularization is omitted, computation steps on the master \eqref{vt}-\eqref{upt} can be removed: each slave reconstructs the image corresponding to its wavelength independently to the others. From a memory point of view, the master node stores the whole dirty image cube and psf, whereas each slave node as mentioned previously only stores the dirty image and PSF at the corresponding band.  Finally, without any loss of generality, a node can be assigned a group of bands, i.e. it can handle a sub-cube of the image rather than just one band. In this case, the communication overhead between the master node and the slave nodes can be reduced compared to the case where each band is assigned to a node.

% Alg. MUFFIN UPDATE 
% ---------------------------------------------------

{\LinesNumberedHidden
\begin{algorithm}
\footnotesize
\begin{align}
\bnabla &= \left [ \bH_1^\dagger(\bH_1\bx_1-\by_1)  | \ldots |  \bH_L^\dagger(\bH_L\bx_L-\by_L)  \right ] \label{grad1} \\  
\btX &= \left( \bX - \tau \left(\bnabla + \bW_s^\dagger \bU\bD_\bmu   + \bD_\bgamma \bV \bW_\lambda \right)   \right)_+  \label{xt1}\\
\bU &=\text{sat}\left( \bU + \sigma  \bW_s (2\btX -\bX)\bD_\bmu  \right), \label{updateu1} \\
\bV &=\text{sat}\left( \bV + \sigma \bD_\bgamma (2\btX -\bX)\bW_\lambda^\dagger  \right)   \label{updatev1} \\
\bX &=  \btX  \label{updatex1} 
\end{align}
\vspace*{-1em}
\caption{MUFFIN update \label{MUFFIN}} 
\end{algorithm}}

% Alg. MUFFIN - Par
% ---------------------------------------------------
\begin{algorithm}
\footnotesize
Each node $\ell=1\ldots L$ computes sequentially:
\begin{align}
\bx_\ell &=  \tilde{\bx}_\ell ,    \label{updatenode2} \\
\boldsymbol{\nabla}_\ell &=\bH_\ell^\dagger(\bH_\ell\bx_\ell-\by_\ell)  \label{grad2} \\  
\bs_\ell &=\mu_\ell \bW_s^\dagger \bu_\ell \label{adjointspat2}  \\
\bm_\ell &= \bx_\ell - \tau(\boldsymbol{\nabla}_\ell+\bs_\ell + \bt_\ell)   \label{stpm}\\
\tilde{\bx}_\ell &= \left( \bm_\ell  \right)_+  \label{stpx} \\
{\bp}_\ell &=\bu_\ell + \sigma\mu_\ell\bW_s(2\tilde{\bx}_\ell -\bx_\ell) \label{stpp} \\
{\bu}_\ell &=\text{sat}\left({\bp}_\ell\right), \label{stpu}
\end{align}

and sends $\bx_\ell$ and $\tilde{\bx}_\ell$ to the master node. 

The master computes sequentially, for $n=1\ldots N$:
\begin{align}
{\tilde{\bv}}^n &=\bv^n + \sigma\gamma_n\bW_\lambda(2\tilde{\bx}^n -\bx^n)   \label{vt} \\
{\bv}^n &=\text{sat}\left( {\tilde{\bv}}^n \right)   \label{v}   \\
\bt^n & = \gamma_n\bW_\lambda^\dagger\bv^n,   \label{upt} 
\end{align}

and sends the col. $\ell$ of $\bT$, denoted as $\bt_\ell$, to node $\ell$. 
\caption{Parallel MUFFIN update \label{MUFFINPar}} 
\end{algorithm}

\section{Risk estimation for optimal parameter selection}

The regularization parameters values $\mu$ and $\gamma$ defined in \eqref{2regpar},  control the strength of the spatial and  spectral regularizations w.r.t. the strength of the data fidelity term. Setting the regularization parameters values appropriately is crucial to ensure a good performance of the deconvolution algorithm, and to obtain a good estimated image $\bX$ of the true sky $\bX^\star$.  However, judging the estimated image quality and its closeness to the true and sought after image is a nontrivial and complicated task. This is due to the lack of ground truth, i.e. due to the fact that the true image is unknown. In addition to this, a `trial and error' technique consisting in the inspection of multiple reconstructions is extremely time consuming when dealing with large-scale problems and multiple parameters. 

The literature on optimal parameter selection, within the frequentist statistical paradigm, provides various quantitative measures for evaluating the quality of an estimated image without prior knowledge of the true image. These measures can be broadly classified as those based on the discrepancy principle \citep{karl2000regularization,morozov1966solution}, the L-curve method \citep{reginska1996regularization,hansen1993use}, the generalized cross-validation (GCV) \citep{golub1979generalized}, and the (generalized) Stein Unbiased Risk Estimator (SURE) \citep{stein1981estimation,eldar2009generalized}.  Among these measures, the SURE is one of the most attractive: unlike the GCV, the L-curve, and the discrepancy principle, the SURE benefits from optimal properties for nonlinear estimators and it is not restricted for linear ones.  It provides an unbiased estimate of the risk i.e. the Mean Square Error (MSE), a quantitative measure often used for quality assessment in image processing.  The SURE measure was originally applicable in the case of image denoising, i.e. when all the $\bH_\ell$ are equal to the identity matrix in equation \eqref{model}.  Nevertheless, the concepts underlying the SURE have been generalized to the case of image reconstruction. Weighted versions of the SURE allow the unbiased estimation of a weighted MSE in the case where the $\bH_\ell$ are not equal to the identity matrix as it is the case in equation \eqref{model}. In this work, we consider a weighted version of the SURE known as the Predicted SURE (PSURE). The PSURE and other weighted SURE metrics have been successfully adopted for automatic tuning in image deconvolution, see for example \citep{ramani2012regularization, giryes2011projected,deledalle2014stein,ammanouil2017multi} and references therein. 

A straightforward strategy for optimal parameter selection based on the SURE criteria or any other metric is by exhaustive search.  More precisely, image reconstructions using all combinations of the regularization parameters are tested to achieve the minimum SURE estimate, see for example the work of  \citet{eldar2009generalized,ramani2008blind}. Compared to the grid search, a golden section allows to narrow the range of values inside which the extremum exists \citep{ramani2012regularization}.  However, the golden section search is most suited for the case of one parameter value. In the case of multiple regularization parameters it can be either used to successively find the optimal regularization parameters values as in  \citep{ammanouil2017multi} or more sophisticated method such as the simplex method can be used. In general, exhaustive search, or bisection strategies for multiple parameters can become rapidly computationally prohibitive.

 More recently, \citet{deledalle2014stein} proposed to minimize the SURE (or PSURE) using a gradient descent approach. This approach is based on an approximation of the SURE that is weakly differentiable w.r.t. the regularization parameters. This approximation, denoted as SURE-FDMC, relies on finite differences and Monte Carlo approximations (FDMC). The weak gradient of the SURE-FDMC, called the Stein Unbiased GrAdient estimator of the Risk (SUGAR) is used in a gradient descent scheme to find the optimal parameters. Among orther advantages, the gradient descent scheme allows setting multiple parameters more easily than grid search and golden section approaches. 

The exhaustive search using  SURE and the gradient descent using SUGAR  all require running the reconstruction algorithm until convergence various times, with different values for the regularization parameters, which can be prohibitive when the algorithm is computationally complex. \citet{giryes2011projected} proposed to update the regularization parameter in parallel to the reconstruction algorithm: at each iteration of the reconstruction algorithm, the regularization parameter is updated by a minimization of the SURE computed a number of iterations ahead using a golden section algorithm. As noticed by \citet{giryes2011projected}, for a given number of iterations of the reconstruction algorithm and golden section, the computational cost of this greedy approach is larger than the global approach. In this work, we propose to estimate the optimal regularization parameters in a more computationally  efficient way. Towards this goal, they are updated in parallel to the reconstruction algorithm using a gradient descent strategy based on SUGAR. More precisely, the regularization parameters are updated every $\Delta$ iterations of the reconstruction algorithm such that the SURE-FDMC is reduced. We refer to this method as ADAptive Parameter Tuning (ADA-PT). To the best of our knowledge, \citep{garsden15,jiang2017joint} are the only works in image reconstruction to propose an `online' self-tuning of the parameters. Their approach relies on the fact that in many sparse reconstruction algorithms based on primal optimization methods, the regularization parameters act at each iteration as thresholds levels on reconstructed images. This is the case in \citep{garsden15} where \citet{Beck2009} FISTA is used with a sparse synthesis prior. The regularization parameters can then be set as a function of a noise level estimation associated to the image update. In contrast with this approach, the application of the ADA-PT approach is not limited to MUFFIN and can be extended to most iterative reconstruction algorithms. 

In what follows, we briefly review the essential concepts behind the estimation of the Predicted Stein Unbiased Risk Estimator based on Finite Differences and Monte Carlo simulations (PSURE-FDMC)  and the Stein Unbiased GrAdient estimator of the Risk introduced by \citet{deledalle2014stein}. We derive the iterative steps necessary for the computation of the SUGAR in the case of MUFFIN. We then show how SUGAR is used to automatically and adaptively select the optimal regularization parameters int ADA-PT. We also show that the steps required for self-tuning can be performed in parallel and in a memory efficient way. 

\subsection{Risk estimation: PSURE-FDMC}

The SURE provides an unbiased estimate of the MSE. However, for ill-posed inverse problems, it is generally not possible to estimate the MSE since the measurements may only contain partial information about the true image. The concepts underlying the SURE have been extended to the case of image reconstruction using weighted versions of the SURE  measure  \citep{ramani2012regularization,giryes2011projected}: the PSURE. The predicted risk, or Predicted Mean Square Error (PMSE) for model \eqref{model} refers to the norm of the discrepancy between the true sky and its estimate in the measurement domain, in other terms it refers to the reconstruction error measured in the $\ell_2$-norm sense: 
\begin{equation}
\mathsf{R} = \sum_{\ell=1}^L \mathsf{R}_\ell, 
\label{prisk0}
\end{equation}
where 
\begin{equation}
\mathsf{R}_\ell = \mathbb{E} \| \bH_\ell(\bx^\star_\ell - \bx_\ell)\|^2.
\label{prisk}
\end{equation}
The PSURE is an unbiased estimator of the PMSE, but unfortunately it is not generally  differentiable, e.g. when considering non-smooth regularizers as in \eqref{RegAll}. For this reason, we will focus from now on on the PSURE-FDMC. The PSURE-FDMC is an asymptotically unbiased estimator for the PMSE in the case of a linear model with additive white Gaussian noise as in \eqref{model}. Let $\hat{\mathsf{R}}$ be the overall PSURE-FDMC. According to \citep{deledalle2014stein}, the PSURE-FDMC is given by: 
\begin{equation}
\hat{\mathsf{R}}  = \sum_{\ell=1}^L \hat{\mathsf{R}}_\ell \\.
\label{PSURE}
\end{equation}
where
\begin{equation}
\label{PSURE0}
\hat{\mathsf{R}}_{\ell} = \| \by_\ell - \bd_\ell\|^2 - N \sigma_w^2 +  \sigma_w^2 \frac{2}{\epsilon} \left< \bod_\ell - \bd_\ell, \bdelta_\ell \right> 
\end{equation}
where  $\bdelta_\ell$ are  independent random Gaussian vector direction such that $\bdelta_\ell \sim \mathcal{N}(0,\bI)$, $\bd_\ell = \bH_\ell \bx_\ell$ and $\bod_\ell = \bH_\ell \bovx_\ell$ with $\bovx_\ell$ the output of the deconvolution algorithm when $\bovy_\ell := \by_\ell + \epsilon \bdelta_\ell$ is used instead of $\by_\ell$ for $\ell=1,\ldots,L$ i.e. when the observations are slightly perturbed, $\epsilon > 0$ is a small constant that should be chosen as small as possible as long as it does not rise to numerical instabilities. Note that $\bdelta_\ell$ does not have to be necessary Gaussian as explained by \citet{avron2011randomized}.
As mentioned in the beginning of this section, the PSURE-FDMC provides an unbiased estimate of the PMSE:
\begin{equation}
\text{lim}_{\epsilon \rightarrow 0} \mathbb{E}\left [ \hat{\mathsf{R}} \right ] = \mathsf{R} . \\
\label{PSURE1}
\end{equation}
The PSURE-FDMC in equation \eqref{PSURE0} only depends on the measurements, the outputs of the deconvolution algorithm ($\bx_\ell$ and $\bovx_\ell$, $l\in 1,\ldots,L$), and requires knowing the noise variances $\sigma_w^2$. Knowledge of the noise variance is generally assumed in interferometric imaging, as in  \citep{Onose16,Purify}, or it can be estimated in the image domain using as in \citep{garsden15} robust statistics such as the median absolute deviation (MAD). Most importantly, note how the PSURE-FDMC in equation \eqref{PSURE0}  does not require the knowledge of $\bx_\ell^\star$, unlike its PMSE counterpart in \eqref{prisk}. As a result, it measures the quality of the estimation for given values of the regularization parameters, and the regularization parameters that gives the best estimate (i.e. the lowest PSURE-FDMC) are the best one among the tested values. 

\subsection{Gradient estimation: SUGAR-FDMC}

Contrarily to the SURE, the PSURE-FDMC defined in \eqref{PSURE0} is differentiable in the weak sense w.r.t. the regularization parameters. Noting that $\bx_\ell$ and $\bovx_\ell$ are the only terms that depend on the regularization parameters, a straightforward derivation of PSURE-FDMC in equation \eqref{PSURE0} w.r.t. one of the regularization parameters yields the SUGAR-FDMC: 
\begin{equation}
\hat{\mathsf{R}}_{\ell}' =  2 ( \bH_\ell \bx_\ell - \by_\ell )^\top \bH_\ell \bx_\ell' +   \frac{2 \sigma_w^2}{\epsilon} \left< \bod_\ell' - \bd_\ell', \bdelta_\ell \right>, 
\label{SUGAR}
\end{equation}
where $\bd_\ell' = \bH_\ell \bx_\ell'$ and $\bod_\ell' = \bH_\ell \bovx_\ell'$, and the notation $(\cdot)'$ refers to the weak derivative w.r.t. one of the regularization parameters, in other terms $(\cdot)'$ corresponds to either $\frac{\partial(\cdot)}{ \partial \mu}$ or $\frac{\partial(\cdot)}{\partial \gamma}$.  Interestingly, it was shown by \citet{deledalle2014stein} that SUGAR-FDMC defined in \eqref{SUGAR} as the weak gradient of PSURE-FDMC, is an unbiased estimate of the weak gradient of the PMSE in \eqref{prisk} when $\epsilon$ tends to zero:
\begin{equation}
\text{lim}_{\epsilon \rightarrow 0} \mathbb{E}\left [ \hat{\mathsf{R}}'  \right ]=  \mathbb{E}\left [ \mathsf{R}'  \right ],
\end{equation}
which means that equation \eqref{SUGAR} gives access to an unbiased estimate of the risk's gradient without prior knowledge of the true sky. 

The computation of SUGAR-FDMC requires $\bx_\ell'$  and $\bovx_\ell'$, the weak derivatives of $\bx_\ell$  and $\bovx_\ell$ respectively, for $\ell=1,\ldots,L$.  Given that $\bx_\ell$  and $\bovx_\ell$ are estimated iteratively, $\bx_\ell'$  and $\bovx_\ell'$ are also estimated iteratively. This is done by simply deriving each equation in algorithm \ref{MUFFINPar}  w.r.t. the regularization parameter under investigation. The iterative algorithms necessary for estimating $\frac{\partial\bx_\ell}{ \partial \gamma}$ and $\frac{\partial \bx_\ell}{\partial \mu}$ are described in algorithms \ref{MUFFINmu} and \ref{MUFFINlambda} respectively. The weak derivatives $\frac{\partial \bovx_\ell}{ \partial \gamma}$ and $\frac{\partial \bovx_\ell}{\partial \mu}$ are estimated by replacing $\by_\ell$ by $\bovy_\ell$ in algorithms \ref{MUFFINmu} and \ref{MUFFINlambda}. Each iteration in algorithm \ref{MUFFINmu} is obtained by deriving the corresponding iteration in algorithm \ref{MUFFINPar}  w.r.t. $\mu$. When a variable implicitly depends on $\mu$, which is the case for $\bx_\ell$ and all other intermediate variables, this consists in simply replacing the variable by its derivative, see for example \eqref{updatenode_mu}. When the variable depends on the product between $\mu$ and another variable (that depends on $\mu$), this requires a simple application of the chain rule, see for example \eqref{adjointspat_mu}. The functions $\mathcal{U}(\cdot)$ and $\Pi(\cdot)$, in equations \eqref{U_mu} and \eqref{pi_mu} respectively, correspond to the weak gradients of the projection on the positive orthant function and the saturation function defined in equations \eqref{projpos} and \eqref{sat} respectively. They are defined as follows: 
\begin{equation}
\mathcal{U}(u) := 
\left\{ \begin{array}{ll}
  0 & \quad \text{if} \quad u \leq 0, \\
  1 & \quad \text{if} \quad u > 0, \\ 
\end{array}\right.
\end{equation}
and 
\begin{equation}
\Pi(u) := 
\left\{ \begin{array}{ll}
  1 & \quad \text{if} \quad -1 \leq u \leq 1, \\
  0 & \quad \text{if} \quad \text{otherwise}, \\ 
\end{array}\right.
\end{equation}
When the input is a matrix, they are applied component-wise. The same concepts are applied for the derivation of algorithm \ref{MUFFINlambda}.

Running $m$ iterations of algorithms \ref{MUFFINmu} and \ref{MUFFINlambda} with algorithm \ref{MUFFINPar} starting from iteration $k$, provides an iterative procedure for estimating the weak derivative of the reconstructed image $m$ steps ahead $k$ w.r.t. $\mu$ and $\gamma$ respectively, and similarly for $\bovx_\ell$ when $\by_\ell$ is replaced by $\bovy_\ell$.  Plugging these derivatives in equation \eqref{SUGAR} provides an asymptotically unbiased estimate of the gradient of the risk $m$ iteration ahead $k$. We will denote the corresponding gradient as $\nabla \hat{\mathsf{R}}_{k:k+m}(\mybtheta)$ where $\mybtheta$ represents either $\gamma$, $\mu$ or $(\gamma, \mu)^\dagger$.

\subsection{Adaptive Parameter selection}

Ideally one would like to find the optimal parameters $\mu^\star$ and $\gamma^\star$ such that the PMSE at the convergence of the MUFFIN algorithm is minimized. An alternative solution for estimating the regularization parameters is to use the PSURE-FDMC  instead of the PMSE:
\begin{equation} 
\mybtheta^\star = \text{argmin}_{\mybtheta}\; \hat{\mathsf{R}},
\label{minR}
\end{equation}
In order to solve problem \eqref{minR} avoiding grid search techniques and for multiple patrameters, the authors of \citep{deledalle2014stein} propose to take advantage of the gradient estimate of the PSURE-FDMC defined in the previous section in order to find the optimal parameters using a gradient descent scheme at convergence of the reconstruction algorithm. The gradient descent approach finds the optimal regularization parameters according to the following updates: 
\begin{equation}
\mybtheta^{(i+1)}  =  \mybtheta^{(i)} - \eta \nabla \hat{\mathsf{R}}_{0:n_c}(\mybtheta^{(i)}) ,
\label{globalgrad}
\end{equation}
where $\eta$ is a sufficiently small step size, $\nabla \hat{\mathsf{R}}_{0:n_c}$ is computed as explained in the previous section and $n_c$ corresponds to the number of iterations required to achieve convergence of MUFFIN. Note that $n_c$ may depend on $\mybtheta^{(i)}$ but can be set without any additional cost monitoring the successive values of $k\mapsto\|\nabla \hat{\mathsf{R}}_{0:k}(\mybtheta^{(i)})\|$. This gradient descent approach requires at iteration $i$ running until convergence MUFFIN with the regularization parameter value $\mybtheta^{(i)}$ and the derivatives $\nabla \hat{\mathsf{R}}$. Afterwards, the parameters $\mybtheta^{(i)}$ are updated to $\mybtheta^{(i+1)}$, MUFFIN and the derivatives estimation algorithms should be run again and so on. 

\begin{figure}
\centerline{\includegraphics[width=.9\columnwidth]{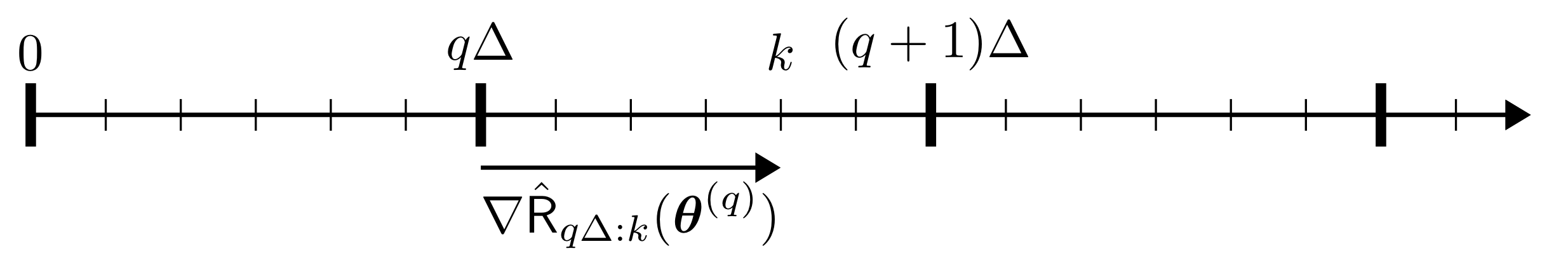}}
\caption{ADA-PT update rule. Thin vertical lines indicate MUFFIN iterations and  thick vertical lines the updates of the regularization parameters.\label{apt_up}}	
\end{figure}

In order to reduce the computational cost of this algorithm, we propose an adaptive gradient descent approach that simultaneously updates the estimates of $\bx^\star_\ell$ and the regularization parameters.  This approach, in the spirit of \citep{giryes2011projected}, was introduced by  \citet{ammanouilEusipco2017} for general iterative algorithms.  More precisely, the regularization parameters are updated every $\Delta$ iterations of MUFFIN according to a gradient descent direction such that the PSURE-FDMC at the current iteration is reduced, see Fig. \ref{apt_up}. At iteration $k$ of MUFFIN, if $k=(q+1)\Delta$, $\btheta^{(q)}$ is updated using
\begin{equation}
\btheta^{(q+1)}  =  \btheta^{(q)} - \eta \nabla \hat{\mathsf{R}}_{q\Delta:k}(\btheta^{(q)}) ,
\label{adaptgrad}
\end{equation}

 Hence, rather than running the MUFFIN algorithm \ref{MUFFINPar} until convergence and then updating the regularization parameters using \eqref{globalgrad}, it requires running algorithms \ref{MUFFINPar}, \ref{MUFFINmu} and \ref{MUFFINlambda} only once. We refer to this method as ADAptive Parameter Tuning (ADA-PT). More precisely, the ADA-PT strategy consists in the iterations described in Algorithm \ref{adapted}. The parameter $\Delta$ sets the trade-off between a greedy algorithm, $\Delta=1$, which will update the parameters at each iteration and the global algorithm \eqref{globalgrad},  $\Delta=n_c$. As pointed by \citet{giryes2011projected}, one of the problems of the greedy strategy ($\Delta=1$) is that it will update the parameters in order to reduce  the estimated PMSE of the current iteration but can harm the overall PMSE.  Thus, instead of choosing  to reduce the estimated PMSE at each iteration, ADA-PT reduces the estimated PMSE every $\Delta>1$ iterations, typically few tens. It is worthy to note that the value of  $\Delta$ does not affect the complexity of the algorithm.
  
 Finally, it is worth pointing out that although estimating the risk gradient entails the additional operations described in algorithms \ref{MUFFINmu} and \ref{MUFFINlambda}, the global memory requirement and computational complexity of ADA-PT'ed MUFFIN are multiplied only by three w.r.t. MUFFIN. Moreover, these additional computations benefit from the same parallel implementation as MUFFIN, described in section \ref{sec:parallel}.

% Alg. d_mu MUFFIN
% ---------------------------------------------------
\begin{algorithm}
\footnotesize
Each node $\ell=1\ldots L$ computes sequentially:
\begin{align}
\bx_\ell' & =  \tilde{\bx}_\ell' ,    \label{updatenode_mu} \\
\boldsymbol{\nabla}'_\ell & =\bH_\ell^\dagger \bH_\ell\bx'_\ell   \\  
\bs_\ell' & = \mu ~  \beta_\ell \bW_s^\dagger \bu_\ell' + \beta_\ell \bW_s^\dagger \bu_\ell \label{adjointspat_mu}  \\
\bm_\ell' & = \bx_\ell' - \tau(\boldsymbol{\nabla}_\ell' +\bs_\ell' + \bt_\ell' )   \\
\tilde{\bx}'_\ell & = \mathcal{U} \left( \bm_\ell  \right) \bm_\ell'   \label{U_mu} \\
{\bp}'_\ell & =\bu'_\ell + \sigma \beta_\ell\bW_s \left(2\tilde{\bx}_\ell -\bx_\ell + \mu (2\tilde{\bx}'_\ell -\bx'_\ell)\right) \\
{\bu}'_\ell & = \Pi\left({\bp}_\ell\right) {\bp}'_\ell,  \label{pi_mu}
\end{align}

and sends $\bx'_\ell$ and $\tilde{\bx}'_\ell$ to the master node. 

The master computes sequentially, for $n=1\ldots N$:
\begin{align}
({\tilde{\bv}}^n)' & = (\bv^n)' + \sigma\gamma \alpha_n\bW_\lambda(2(\tilde{\bx}^n)' -(\bx^n)')    \\
({\bv}^n)' & = \Pi\left( {\tilde{\bv}}^n \right) ({\tilde{\bv}}^n)'    \\
(\bt^n)' & = \gamma \alpha_n\bW_\lambda^\dagger (\bv^n)',  
\end{align}

and sends the col. $\ell$ of $\bT'$, denoted as $\bt_\ell'$, to node $\ell$. 

\caption{Parallel update of  $\bx_\ell' = \frac{ \partial \bx_\ell}{\partial \mu } $ \label{MUFFINmu}} 
\end{algorithm}

% Alg. d_lambda MUFFIN
% ---------------------------------------------------
\begin{algorithm}

\footnotesize
Each node $\ell=1\ldots L$ computes sequentially:
\begin{align}
\bx_\ell' & =  \tilde{\bx}_\ell' ,    \label{updatenode} \\
\boldsymbol{\nabla}'_\ell & =\bH_\ell^\dagger \bH_\ell\bx'_\ell   \label{grad} \\  
\bs_\ell' & = \mu ~  \beta_\ell \bW_s^\dagger \bu_\ell'  \label{adjointspat}  \\
\bm_\ell' & = \bx_\ell' - \tau(\boldsymbol{\nabla}_\ell' +\bs_\ell' + \bt_\ell' )   \\
\tilde{\bx}'_\ell & = \mathcal{U} \left( \bm_\ell  \right) \bm_\ell'  \\
{\bp}'_\ell & =\bu'_\ell + \sigma \mu \beta_\ell\bW_s \left(2\tilde{\bx}'_\ell -\bx'_\ell \right) \\
{\bu}'_\ell & = \Pi\left({\bp}_\ell\right) {\bp}'_\ell, \label{updateu}
\end{align}

and sends $\bx'_\ell$ and $\tilde{\bx}'_\ell$ to the master node. 

The master computes sequentially, for $n=1\ldots N$:
\begin{align}
({\tilde{\bv}}^n)' & = (\bv^n)' + \sigma \alpha_n\bW_\lambda(2\tilde{\bx}^n -\bx^n + \gamma (2(\tilde{\bx}^n)' -(\bx^n)'))   \\
({\bv}^n)' & = \Pi\left( {\tilde{\bv}}^n \right) ({\tilde{\bv}}^n)'  \\
(\bt^n)' & = \gamma \alpha_n\bW_\lambda^\dagger (\bv^n)' + \alpha_n\bW_\lambda^\dagger (\bv^n),  
\end{align}

and sends the col. $\ell$ of $\bT'$, denoted as $\bt_\ell'$, to node $\ell$. 

\caption{Parallel update of $\bx_\ell' = \frac{\partial \bx_\ell}{ \partial \gamma} $ \label{MUFFINlambda}} 
\end{algorithm}

\begin{algorithm}
\begin{algorithmic}[1]
\STATE Set $k=0$ and $q=0$. Initialize $\mybtheta^{(0)}$, $\bX^{(0)}$ and $\overline{\bY}=\bY +\epsilon \boldsymbol{\delta}$. 
\STATE 1 iteration of alg. \ref{MUFFINPar} using $\btheta^{(q)}$ and $\bY$: $\bX^{(k+1)}$
\STATE 1 iteration of alg. \ref{MUFFINPar}  using $\btheta^{(q)}$ and $\overline{\bY}$: $\overline{\bX}^{(k+1)}$
\STATE 1 iteration of algs. \ref{MUFFINmu} and \ref{MUFFINlambda} for $\bX^{(k+1)}$
\STATE 1 iteration of algs. \ref{MUFFINmu} and \ref{MUFFINlambda} for $\overline{\bX}^{(k+1)}$
\IF{$k = (q+1)\Delta$} 
	\STATE {Compute $\nabla \hat{\mathsf{R}}_{q\Delta:k}(\mybtheta^{(k )}) $ using \eqref{SUGAR}.} 
	\STATE {Update  $\mybtheta^{q}$ to $\mybtheta^{q+1}$ using \eqref{adaptgrad}.}
	\STATE $q \leftarrow q+1$
\ENDIF
\STATE $k \leftarrow k+1$. Go to 1:
\end{algorithmic}
\caption{ADA-PT'ed MUFFIN algorithm}
\label{adapted}
\end{algorithm}

\section{Tests on simulated data}

The reconstruction results reported in this section were obtained with the proposed distributed implementation of MUFFIN. The aim of these simulations is twofold: 1) to demonstrate  the efficiency of the distributed implementation; 2) to illustrate the relevance of the ADA-PT strategy for automatically finding the optimal regularization parameters. 

\subsection{Data generation description}

The  data sets used in the following are composed by two simulated, multi-wavelength radio sky emissions in which we have  introduced complex spatial and spectral features in order to challenge the reconstruction algorithms. The raw data are obtained by   convolving these sky emissions with wavelength dependent PSFs and adding noise. We describe in more details the considered sky emissions and the PSFs  below. Note that in all Figures the flux scale is arbitrary as we are mostly interested in comparing the relative intensities of reference and reconstructed images/cubes. Point estimation of the Signal to Noise Ratio (SNR) and of the PMSE denoted as WMSE: 
\begin{align}
&\text{SNR}(\bX) = 	10\log_{10}\left(\frac{\sum_{\ell = 1}^L \|\bx^\ast_\ell\|_2^2}{\sum_{\ell = 1}^L \|\bx^\ast_\ell - \bx_\ell \|_2^2}\right)\\
&\text{WMSE}(\bX) = \frac{1}{NL}\sum_{\ell = 1}^L \|\bH_\ell(\bx^\ast_\ell - \bx_\ell) \|_2^2
\end{align} 
will however be indicated. 

\subsubsection{Simulated galaxy with H{\small I} emission line (`Galaxy-HI')}

\begin{figure}
\centerline{
   \includegraphics[height=0.35\columnwidth]{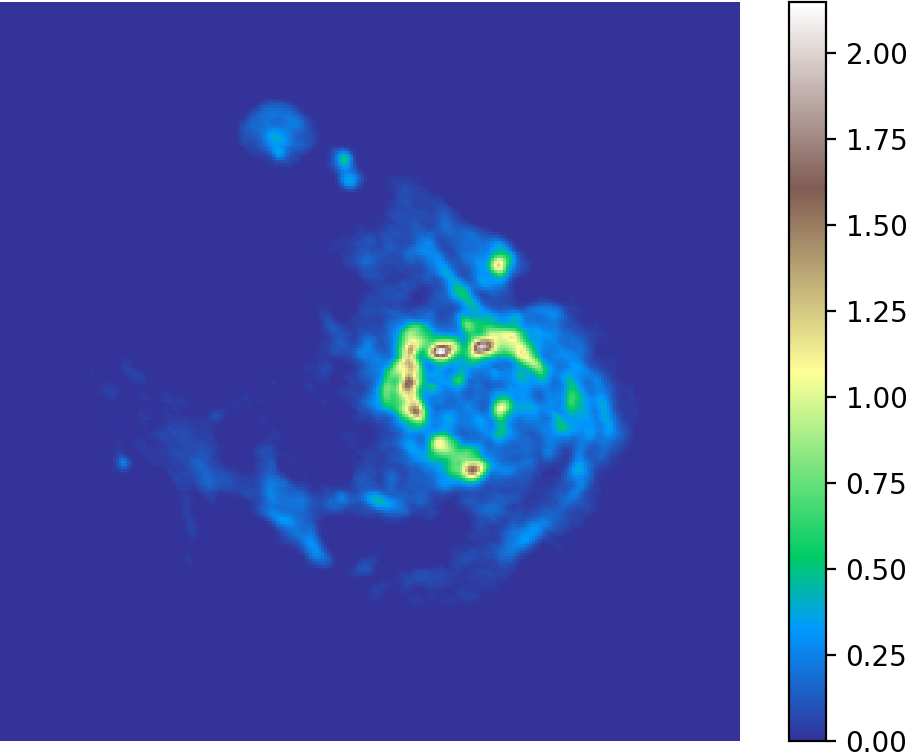}\hfill
   \includegraphics[height=0.35\columnwidth]{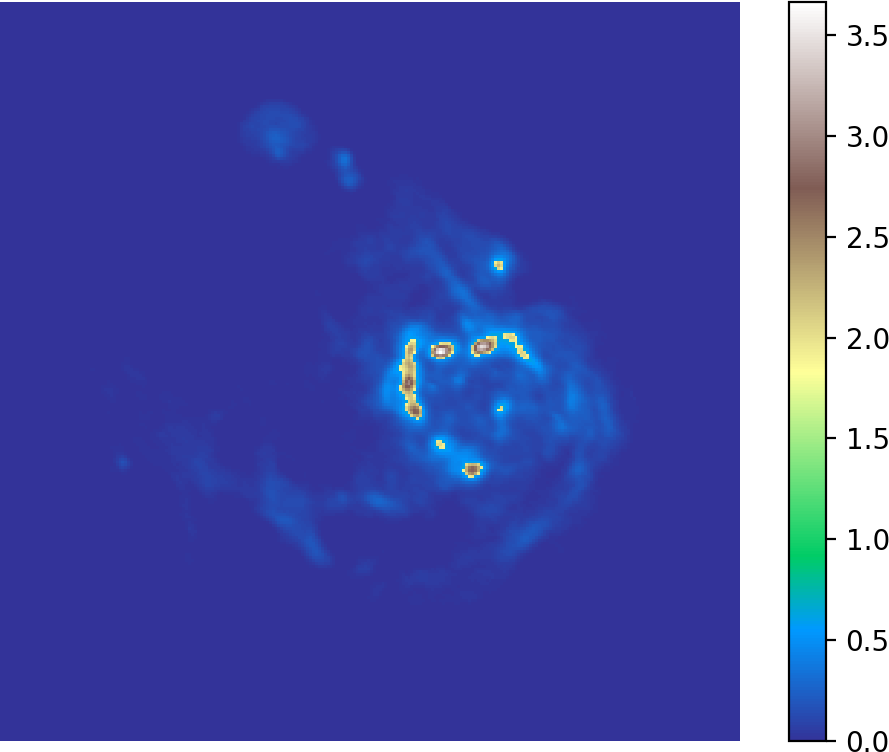}
  }
  \medskip
 \centerline{
   \includegraphics[height=0.35\columnwidth]{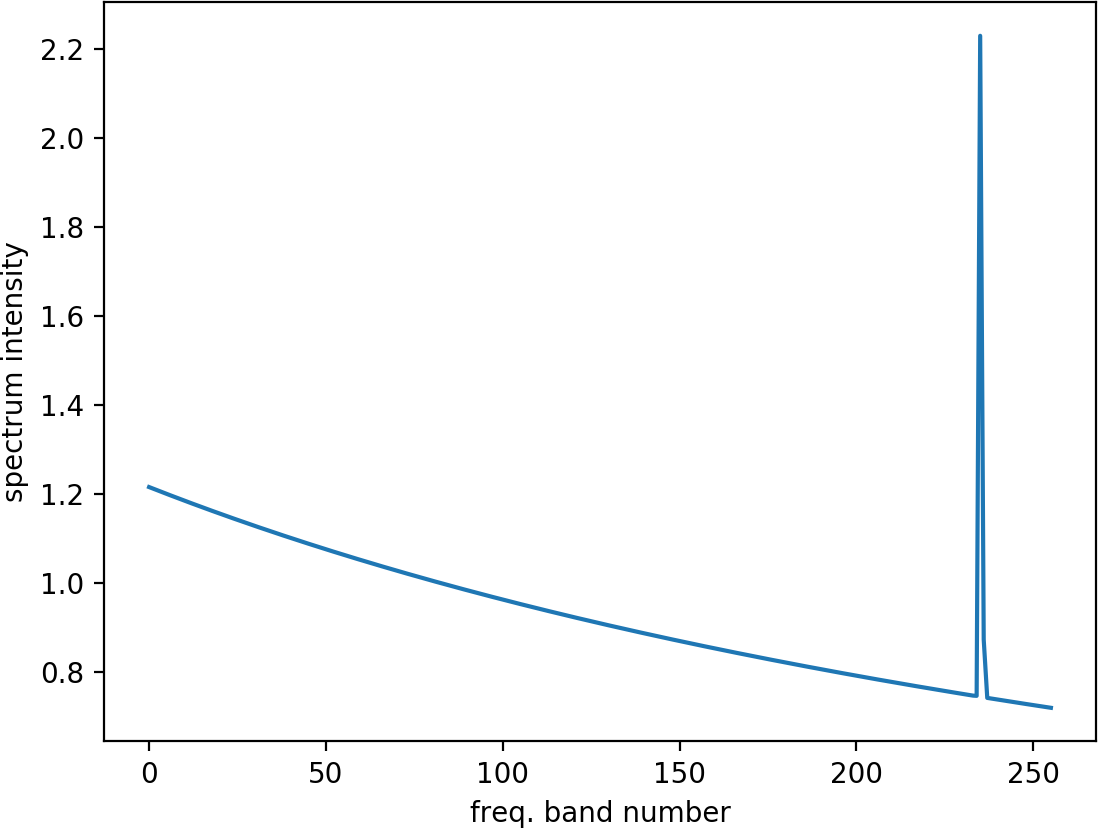}
 }
\caption{`Galaxy-H{\small I}' data set. Top left panel: considered galaxy data image in the first wavelength channel. The H{\small I} emission is added to the brightest pixels (in red, above $1.6$ in this colorscale). Top right panel: image in the H{\small I} emission line ($236$-th wavelength band). Bottom panel: the spectrum of a bright pixel. The intensity peak caused by the H{\small I} emission is visible in the $235^{\text{th}}$ frequency band. The shape of the continuum is a power law.}
\label{fig:GalaxyImage}
\end{figure}

The first data set is a RA-Dec-$\nu$ cube generated from a well known image representing a star-forming region in the M$31$ galaxy (see Fig. \ref{fig:GalaxyImage}, top left panel).  For each of the $256\times 256$ pixels in this image,  we computed a spectrum in $256$ adjacent frequency bands of constant bandwidth  ($2\, \text{MHz}$), starting from $950\, \text{MHz}$. Each spectrum follows a first order power-law spectrum model depending on a spectral index.  The $256\times 256$ map of spectral indices was constructed following the procedure detailed by \citet{Junklewitz2015}: for each pixel, the spectral index is taken as a linear combination of a homogeneous Gaussian field and the reference sky image.  

As an additional spectral feature, we inserted in the spectra of the brightest pixels (see legend of Fig. \ref{fig:GalaxyImage}) a H{\small I} line simulating the localized presence of neutral hydrogen gas. The line is centered at $1.42$ GHz, corresponding to our $235^{\textrm{th}}$ channel. We assume here a simple model where the galaxy has a gas consumption time scale of typically 1 Gyr, a H mass of  $10^9$ solar masses and thus a star formation rate of one solar mass per year. Assuming further a velocity dispersion of $200$ km/s, we obtain lines whose flux is about twice the flux of the continuum at $1.42$ GHz, a value that may typically fluctuate within one  order of magnitude (P. Serra, private communication). Translated into frequency, the velocity dispersion leads to a line width of 1 MHz, with  tails that slightly smear out on adjacent channels. In view of these elements,   the spectral H{\small I}  line profile was computed as a continuous Gaussian emission line profile with a maximum corresponding to twice the local continuum and a FWHM of 1 MHz. The profile was further convolved by a boxcar function of width $2\, \text{MHz}$, sampled at the locations of the considered $256$ frequency bands and added to the continuum. The top right  panel of Figure \ref{fig:GalaxyImage} shows the simulated galaxy in the channel where the H{\small I} line has most energy  ($235^{\text{th}}$ wavelength band). Figure \ref{fig:GalaxyImage} bottom panel shows the spectrum of a bright pixel from the Galaxy data set where the H{\small I} emission was added. We refer to this simulated sky data cube as `Galaxy-HI' for short.

\subsubsection{Simulated galaxy cluster with diffuse halo (`Halo')}

The second data set is a  galaxy cluster radio model based on a  refined version of 2D radio simulations considered in our previous works \citep{CFerrari2015}. This model includes various kinds of radio sources that can be found in  `radio-loud' clusters, in particular unresolved point-like objects, bright elongated radio galaxies with complex morphologies and  a giant low-surface brightness source of synchrotron radiation (or `radio halo'), see e.g. \citep{CFerrari2008,Feretti12} and references therein.  In addition to these features, this simulation  also properly models the number counts of both the cluster radio source and the foregound/background population. The simulated radio halo has been produced following the approach described in \citet{murgia04} while discrete radio sources have been modelled according to what described in \citet{loi18}

To generate the 3D data cube, we segmented the simulated 2D radio map described above into three contributions:  two giant radio galaxies (with the typical head-tail and narrow-angle-tail morphology of FRI in clusters) and a low-surface brightness extended halo. Each of the three contribution was given a specific power law spectrum $\be_k$ ($k=1,2$ or $3$) that remains  spatially constant  up to a multiplicative factor $\alpha_n(k)$.  Formally, spectrum $n$ of the blended data cube  is thus generated as:
$$\bx^{\star\,n} = \sum_{k=1}^3 \alpha_n(k)\be_k,\; \forall n=1,\ldots, N.$$ 
The three considered spectra $\{\be_k\}$,  also called endmembers, are shown in the left panel of Figure \ref{fig:HaloEndAbd}. The three considered segmentation maps (also called abundance maps),  which show in each pixel the fraction $\alpha_n(k)$ of the $k^{\textrm{th}}$ contribution,  are displayed in the three other panels of Figure \ref{fig:HaloEndAbd}. Note that the halo is normalized such as  its power equals  approximately 11 \%  of the power of the final cube.

The purpose of this data set  is twofold. First, to evaluate the ability of the proposed method to reconstruct scenes with very high dynamic range and complex structures (the halo in particular), which is one of SKA first objectives. Second, to show that,  thanks to MUFFIN's non parametric approach, the reconstructed cube can be further processed to achieve a specific task such as the deblending of the sources using their spectral properties. For this spectral unmixing case, a two-step procedure (reconstruction and calibration within a major/minor loop followed by the deblending), compared to a joint procedure such as in \citep{jiang2017joint,abdulaziz2016low}, can be advocated since i. it avoids the resolution of global non convex optimization problems and ii. it gives some space to a human expertise during the unmixing task, e.g. for identifying the endmembers when complex objects are superposed to a halo. 

The final cube, which will be referred to as `Halo' for short, spans a frequency range from 950 MHz to 1.458 GHz, with 256 spectral channels of equal bandwidth.

\subsubsection{Point spread function (PSF)}

\begin{figure}
  \centerline{
   \includegraphics[height=0.35\columnwidth]{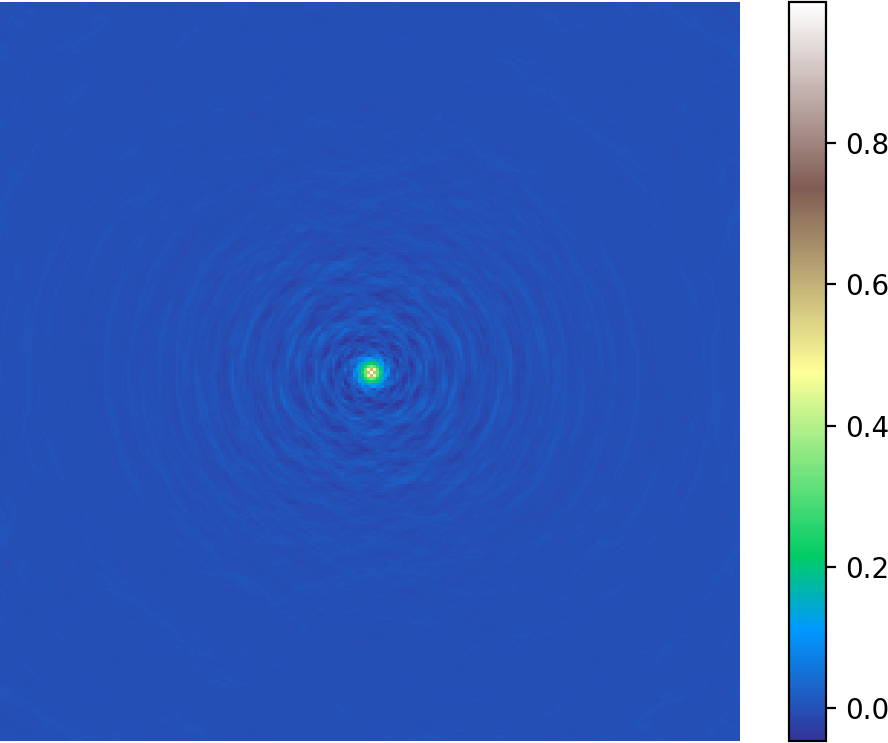}\hfill
   \includegraphics[height=0.35\columnwidth]{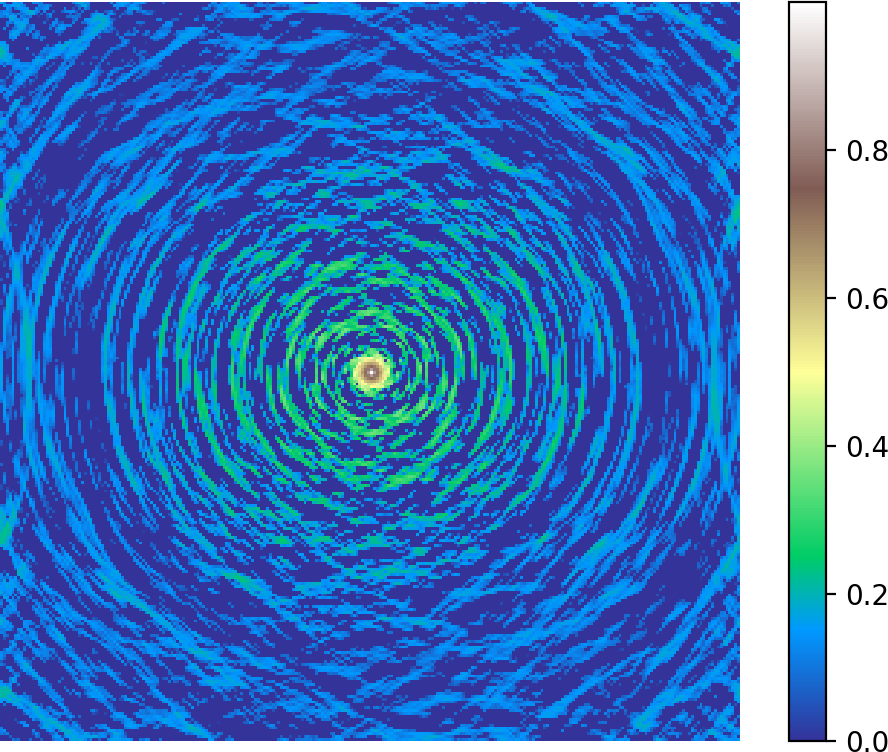}
  }
\caption{Point Spread Function (PSF) in the first channel in linear scale (left) and power  scale ($0.3$, right). The PSF has a central lobe and  ringed side-lobes extending at large angular distances.}
\label{fig:psf}
\end{figure}

We simulated the PSFs of SKA1-MID with its 197 dishes using the HI-inator package\footnote{\url{https://github.com/SpheMakh/HI-Inator}} based on the MeqTrees software by \citet{noordam2010meqtrees}. The simulated PSFs correspond to a Fourier coverage produced by a total observation time of $8$ hours. Figure \ref{fig:psf} shows the PSF at the first wavelength band. The PSF in each channel is characterized by a  central lobe and ringed side-lobes at larger angular distances. This cube of PSFs was used to convolve channel by channel the Galaxy-H{\small I} and the Halo data sets. Gaussian noise was added to the result to produce the dirty data cubes with an SNR of 20 dB for the dirty Galaxy-HI data set and an SNR of 40 dB for the dirty Halo data set.

%671929%%%%%%%%%%%%%%%%%% EX %%%%%%%%%%%%%%%%%%%
\subsection{Deconvolution examples}

\subsubsection{Implementation details}

We tested the proposed deconvolution approach, namely ADA-PT'ed MUFFIN, on the two data sets described above. In all the simulations, $\textbf{W}_s$ and $\textbf{W}_\lambda$ were respectively set to the union of the first $8$ Daubechies wavelet bases and to the Discrete Cosine Transform (DCT). The parameter of the deconvolution algorithm $\sigma$ was set to $\sigma=10$ and the corresponding value of $\tau$ was calculated as explained in section \ref{sec:optimalg}. The parameter $\Delta$ was set to $10$. 

We compare the results of the proposed approach with the MS-MFS (Multi-Scale Multi-Frequency) deconvolution algorithm. We used the implementation developed within the SKA Science Data Processor Consortium available online\footnote{\url{https://github.com/SKA-ScienceDataProcessor/algorithm-reference-library}} with the default parameters. In particular, the number of iterations is equal to $10^3$, the scales are set to $(0, 3, 10,30)$ and order 3 polynomials are used to fit the spectra. We tested MS-MFS using the default parameters and using other parameters, the default parameters gave the best results. 

This work was granted access to  HPC ressources. In particular, the simulations were performed using $80$ compute nodes, 1 master node and 79 slave nodes, with $64$ Go of RAM per node.  This setting  corresponds to $4$ spectral bands at most per slave node. The algorithm is written in Python, and the Message Passing Interface (MPI) is used to deal with the parallel computing architecture.  The computing time to process a $256 \times 256 \times 256$ cube is: 3  sec./iteration for MUFFIN and 15 sec./iteration for the full ADP-PT'ed code. The code is available at \url{https://github.com/andferrari/MUFFIN}.

The most computationally intensive steps of  MUFFIN are the $L=256$ spatial wavelets decompositions and recompositions that are performed at each iteration. Nevertheless, these steps are performed in parallel, see Eq. (\ref{adjointspat2}, \ref{stpp}) in Alg. \ref{MUFFINPar}. On the other hand, MS-MFS requires 12 scale decompositions (4 scales and order 3 polynomials) that are performed at the beginning of the algorithm, i.e. before the first iteration. Memory wise, the amount of memory required by MUFFIN is to store the matrices $\bU$, $\bX$, $\tilde{\bX}$ and $\bV$. As a result, the global memory footprint of MUFFIN mainly is proportional to the total size of these matrices which is equal to  $11LN$. Note that the memory load is spread across the $L$ computing nodes. On the other hand, the MS-MFS memory usage is $12N$. The memory requirements for MUFFIN are doubled for ADA-PT'ed MUFFIN.

\subsubsection{Galaxy-H{\small I} data set}

\begin{figure}
 \centerline{
   \includegraphics[width=0.5\columnwidth]{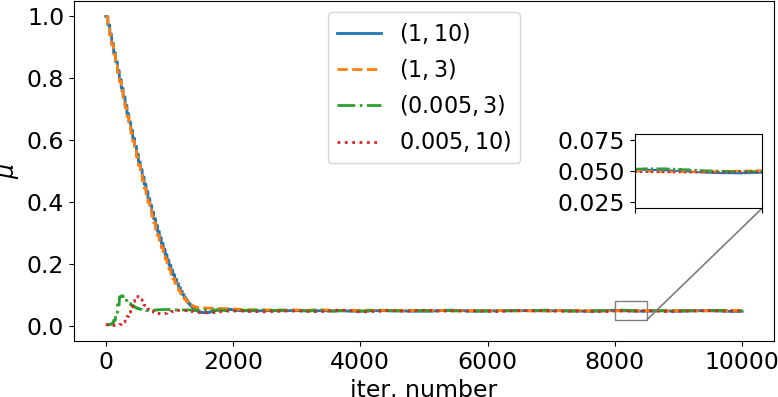}\hfill
    \includegraphics[width=0.5\columnwidth]{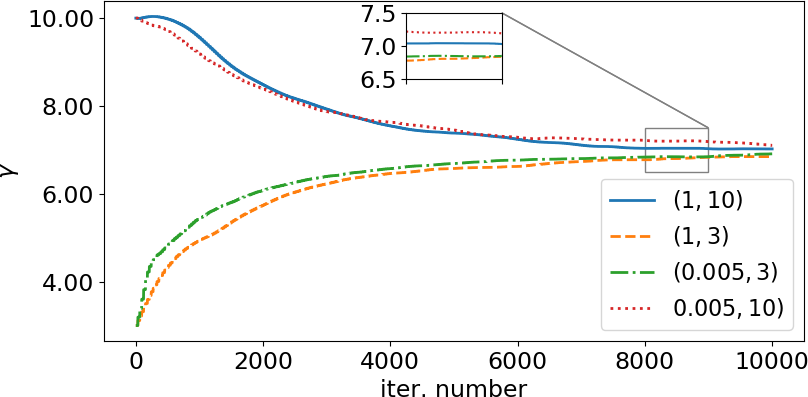}
  }

\caption{Evolution of  $\mu$ (left) and $\gamma$ (right) as a function of the iteration number obtained with four different initialization settings for the regularization parameters $(\mu,\gamma)$: $(1,10)$, $(1,3)$, $(0,005,3)$ and $(0.005,10)$. }
\label{fig:regvar}
\end{figure}

The spatial and spectral regularization parameters $(\mu,\gamma)$  were initialized to four different couples of values: $(1,10)$, $(1,3)$, $(0,005,3)$ and $(0.005,10)$.  The ADA-PT'ed MUFFIN algorithm was ran for $10^4$ iterations and Figure \ref{fig:regvar} shows the evolution of each  regularization parameters for the four different initialization settings as a function of the iterations.  In all cases, the  regularization parameter $\mu$  converges to $0.05$ and the regularization parameter $\gamma$ converges to $7$ (on average). Note that the four initialization settings cover the fours distinct scenari where both parameters values are greater than the optimal ones, less than the optimal ones, and where one is greater and one is less than the optimal value.  
 
\begin{figure}
  \centerline{
   \includegraphics[width=0.5\columnwidth]{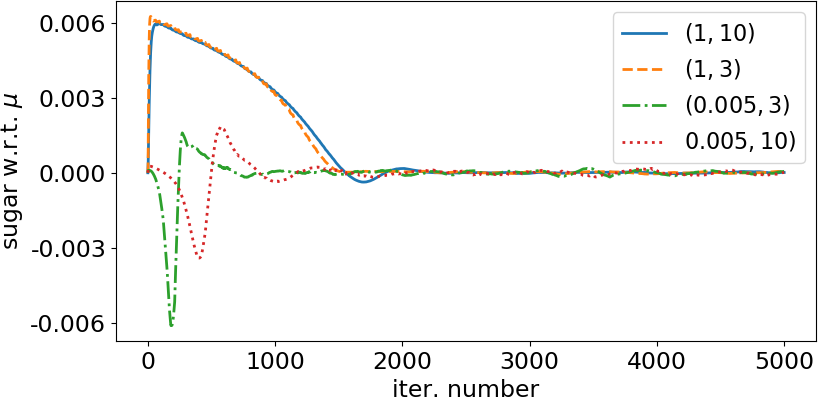}\hfill
   \includegraphics[width=0.5\columnwidth]{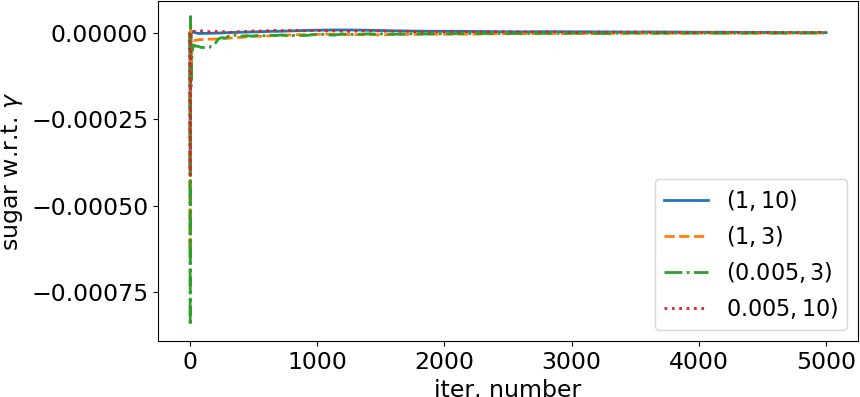}
  }
\caption{Variation of  SUGAR w.r.t. $\mu$ and of SUGAR w.r.t. $\gamma$ as a function of the iteration number for the four considered initialization settings.} 
\label{fig:sugarvar}
\end{figure}

\begin{figure}
  \centerline{
   \includegraphics[width=0.5\columnwidth]{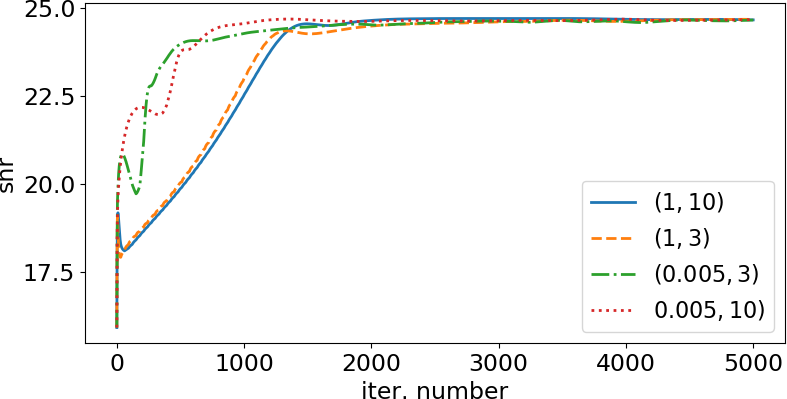} \hfill
 \includegraphics[width=0.5\columnwidth]{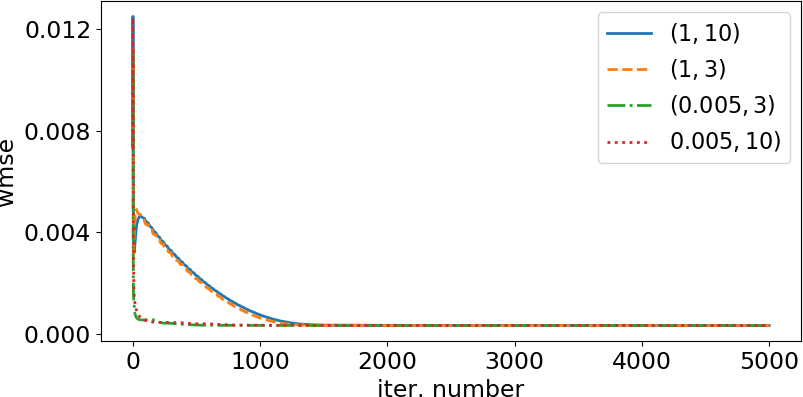} 
  }
\caption{Variation of  SNR and WMSE as a function of the iteration number obtained for the four considered initialization settings.} 
\label{fig:wmsevar}
\end{figure}

Figure \ref{fig:sugarvar} shows the evolution of $k\rightarrow \nabla \hat{\mathsf{R}}_{q\Delta:k}(\mybtheta^{(k )})$, i.e. of SUGAR w.r.t. $\mu$ and of SUGAR w.r.t. $\gamma$ as a function of the iteration number for the four considered  initializations values. Note how both SUGAR curves converge to zero, which justifies the regularization parameters convergence.  Figure \ref{fig:wmsevar} shows the variation of  SNR and  WMSE as a function of the iteration number for the four considered initialization cases. The SNR converges to a value of 24 dB and the WMSE converges to $3.4 \times 10^{-4}$. Note that the SNR and WMSE curves are only shown for the first $5000$ iterations as they remain stable after the first $2000$ iterations.  

\begin{figure}
  \centerline{
   \includegraphics[height=0.4\columnwidth]{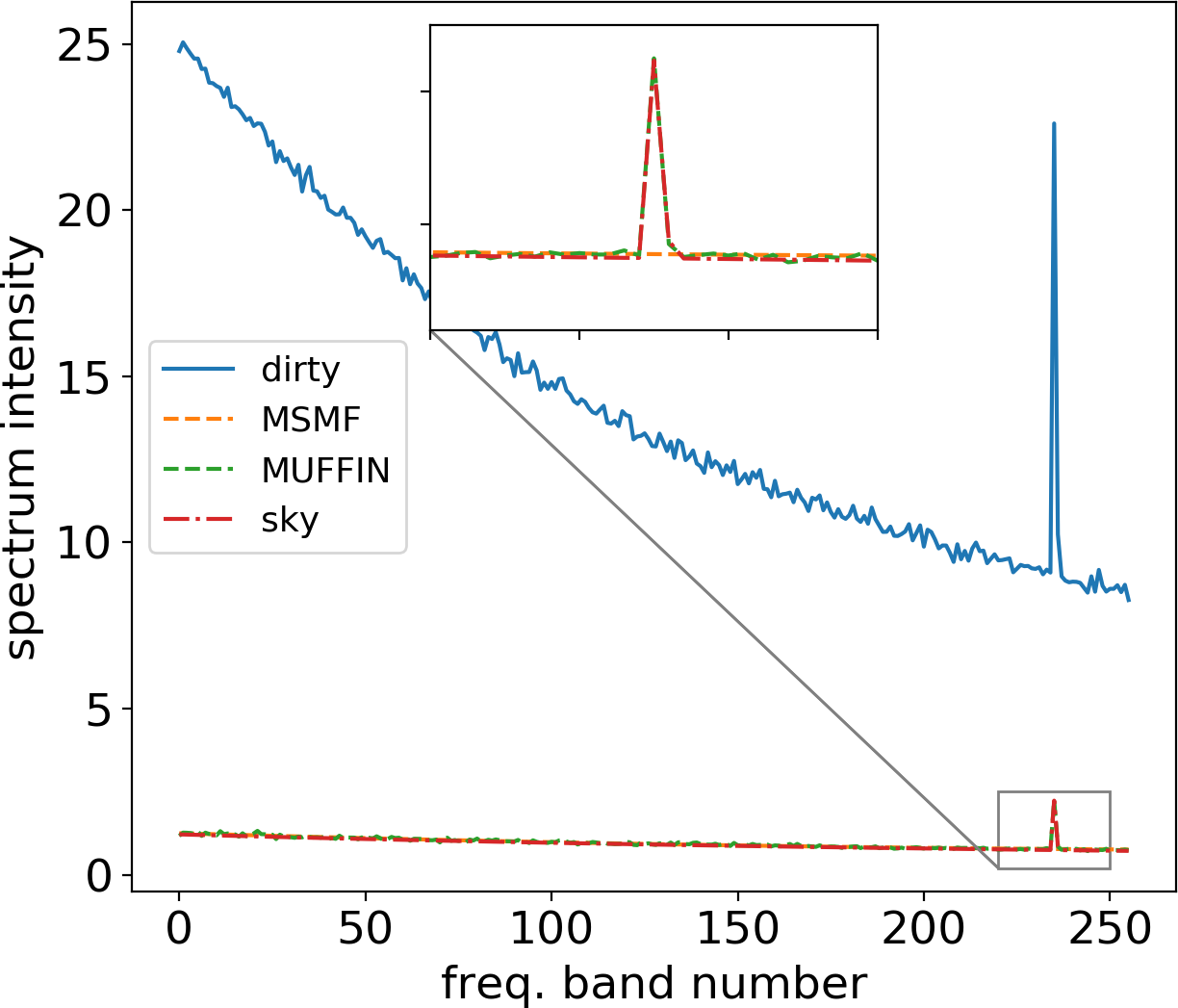}
  }
\caption{Reference, dirty and reconstructed spectra for the bright pixel  characterized by an HI emission shown in Fig. \ref{fig:GalaxyImage}.} 
\label{fig:galaxyresspec}
\end{figure}

\begin{figure*}

  \centerline{
   \includegraphics[height=0.35\columnwidth]{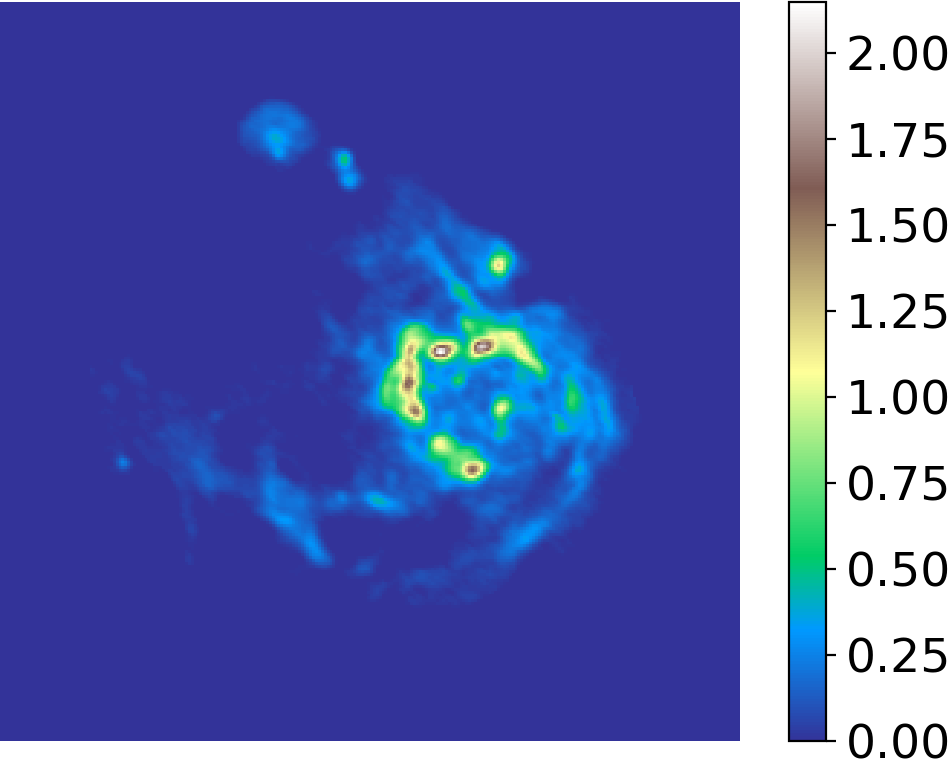} \hfill
   \includegraphics[height=0.35\columnwidth]{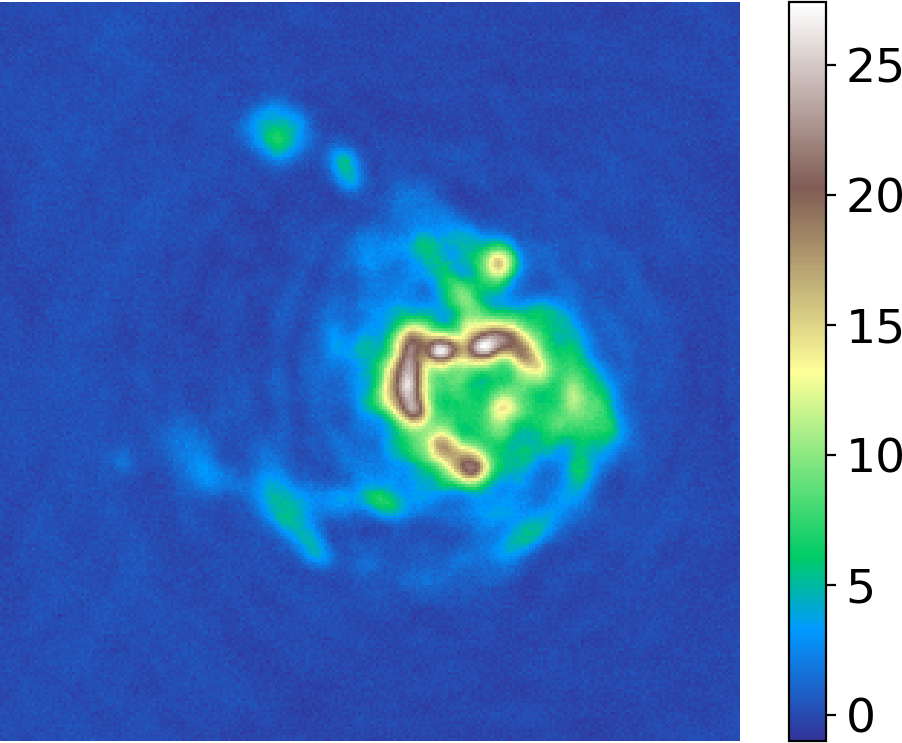} \hfill
   \includegraphics[height=0.35\columnwidth]{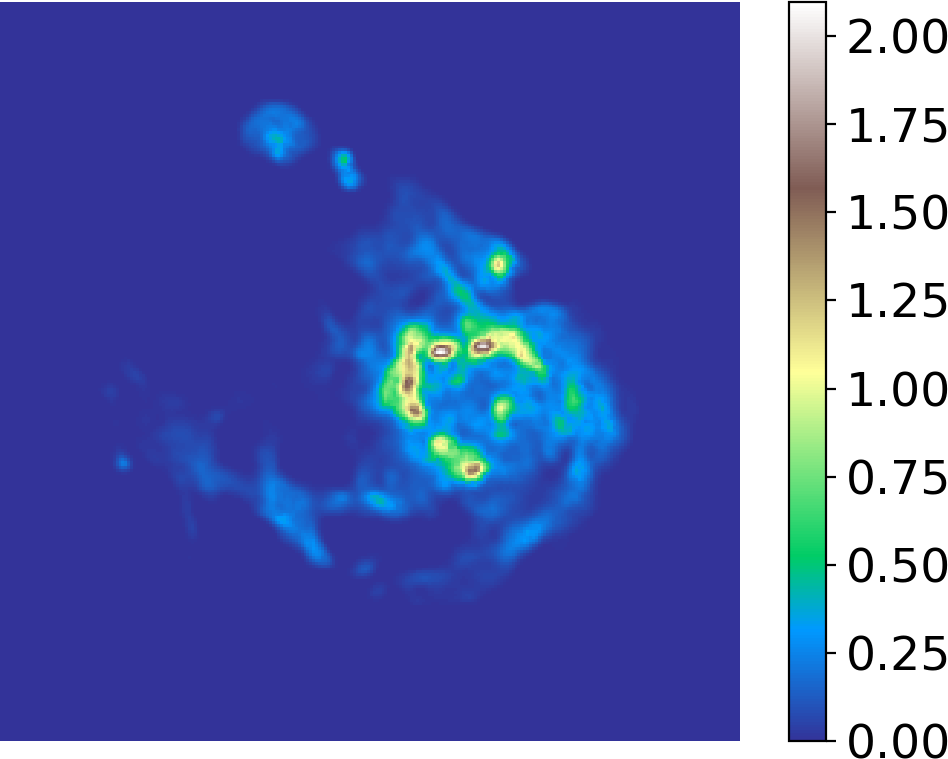} \hfill
   \includegraphics[height=0.35\columnwidth]{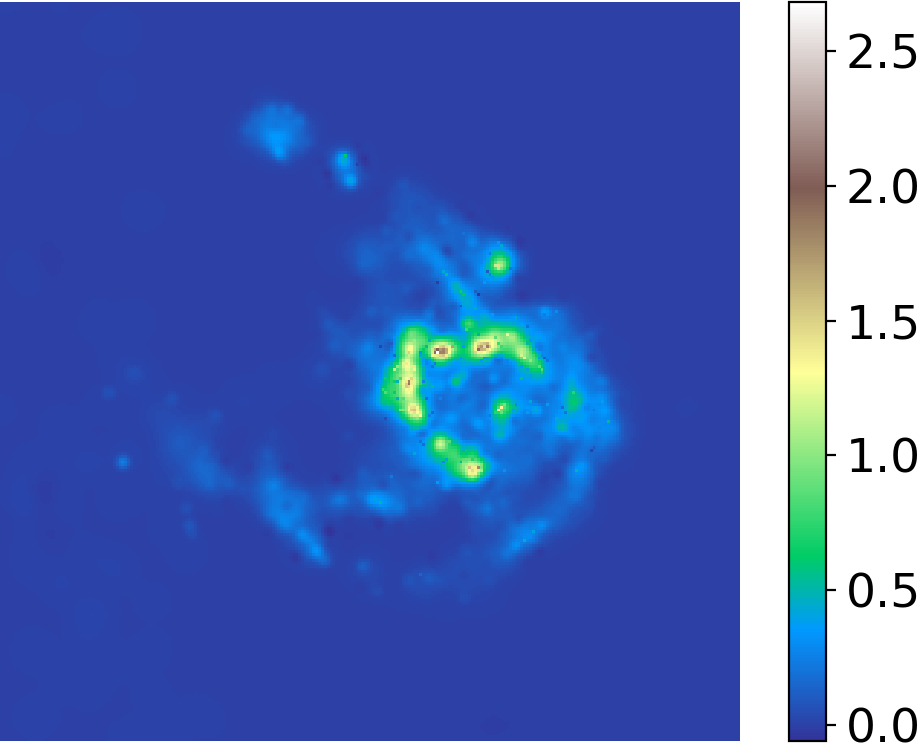}
  }
  \medskip 
  \centerline{
   \includegraphics[height=0.35\columnwidth]{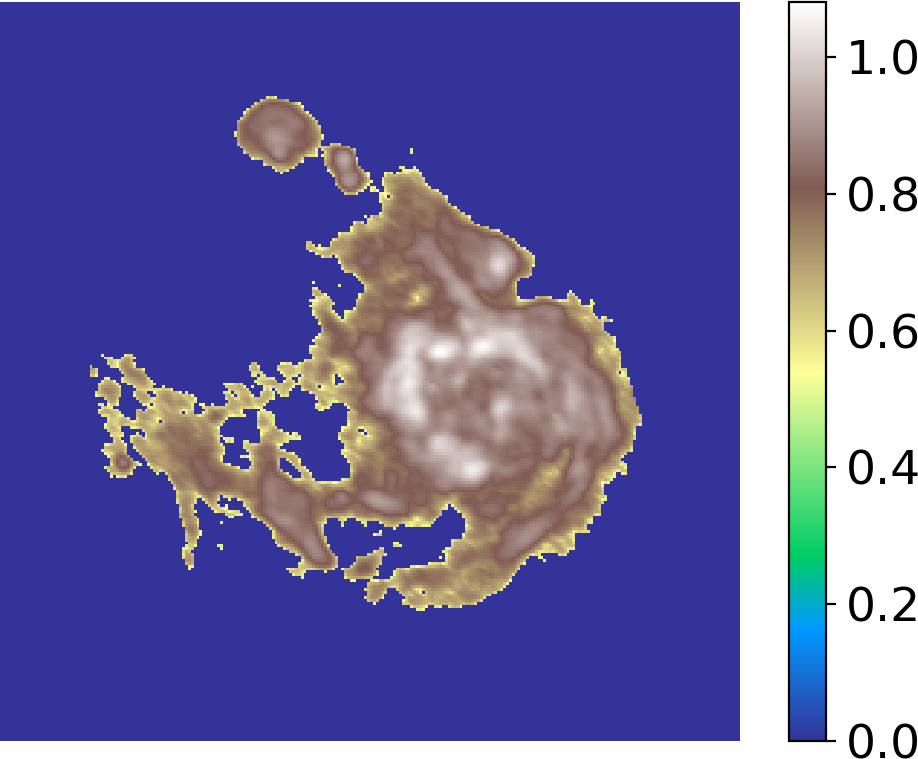} \hfill
   \includegraphics[height=0.35\columnwidth]{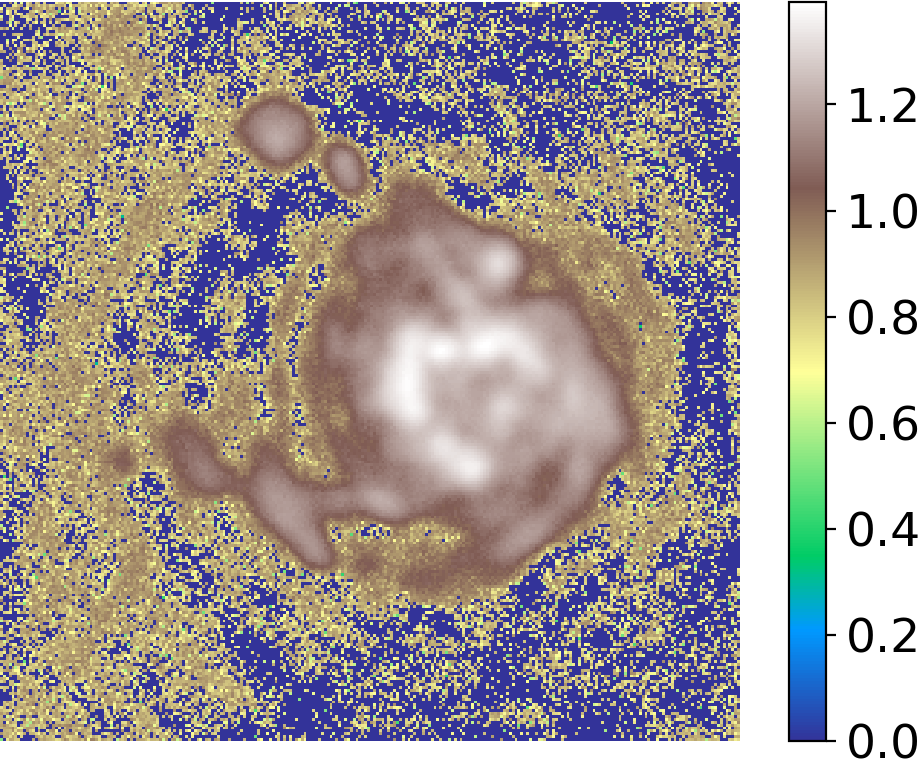} \hfill
   \includegraphics[height=0.35\columnwidth]{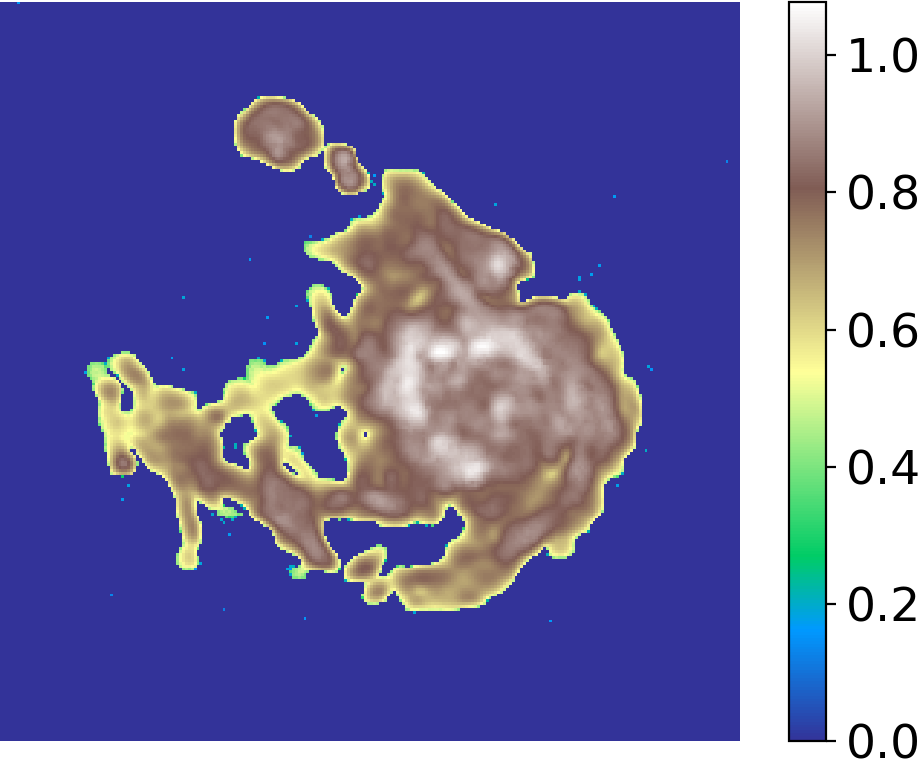} \hfill
   \includegraphics[height=0.35\columnwidth]{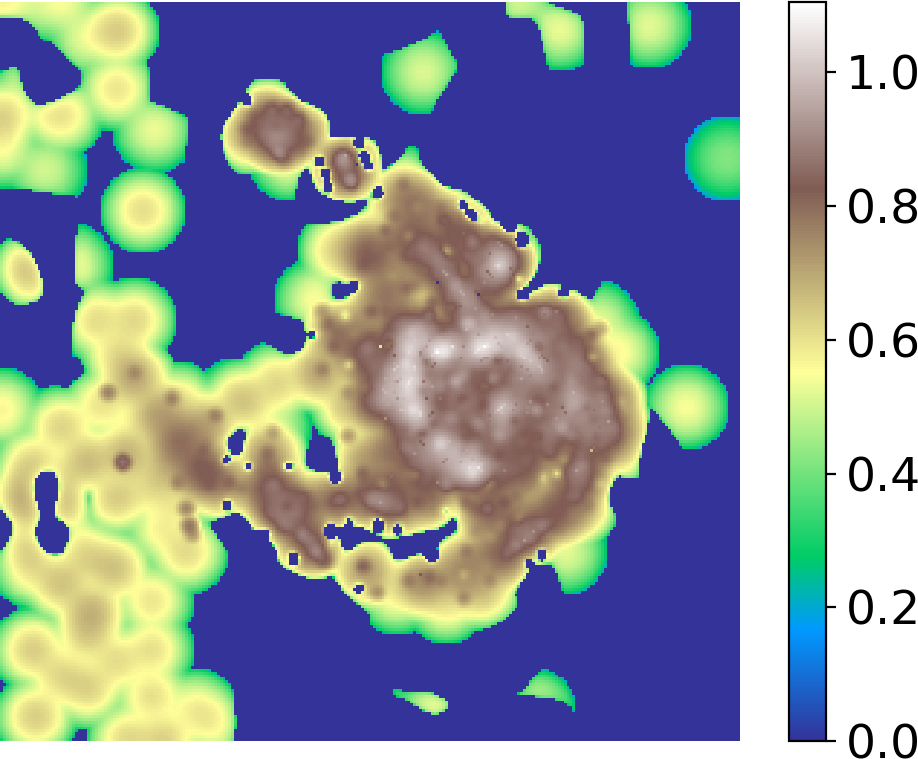}
  }

\caption{First column: Galaxy-H{\small I} reference image. Second column: dirty image. Third column:  reconstructed image using MUFFIN. Fourth column:  reconstructed image using MS-MFS. All images are shown at the first frequency band. First row shows the images intensities at linear scale while the second row shows the images intensities to the power $0.1$ in order to better highlight the small intensities.}
\label{fig:galaxyresim}
\end{figure*}

We compare in Figure \ref{fig:galaxyresim} the proposed approach with \citet{rau2011multi} MS-MFS. The third and fourth column in the first row of Figure \ref{fig:galaxyresim} show that the deconvolved image obtained using MUFFIN and MS-MFS  are both very close to the sky image and `clean' when compared to the dirty image shown in the second column. The second row of Figure \ref{fig:galaxyresim}, which highlights the low intensities,  reveals that MUFFIN was able to better reconstruct the image while some smooth artifacts are seen in the MS-MFS image. Note that the SNR of the reconstructed image cube is $24.6$ dB for MUFFIN and $18.8$ for MS-MFS. Finally, Figure \ref{fig:galaxyresspec} shows the reconstruction results for the spectrum of the bright pixel in the image characterized by an H{\small I} emission and shown in Fig.  \ref{fig:GalaxyImage}.  It can be seen  that both MUFFIN and MS-MFS are able to estimate and denoise the continuum part of the spectrum.  Turning to the inset panel (zoom on the H{\small I} emission line), we see that the line is logically absent in the MS-MFS spectrum, because this algorithm is aimed at reconstructing a power law spectrum. In contrast, MUFFIN is able to preserve the H{\small I} emission in the reconstructed spectra. 

\subsubsection{Halo data set}

The proposed ADA-PT strategy  was run  on  the Halo data set, for different initialization settings and for $10^4$ iterations. The  regularization parameter $\mu$ and $\gamma$  converge respectively to $7 \times 10^{-6}$ and  to $2.6 \times 10^{-2}$. The SNR converges to a value of 31 dB, and the WMSE converges to $1.6 \times 10^{-7}$. 

\begin{figure*}
  \centerline{
   \includegraphics[height=0.35\columnwidth]{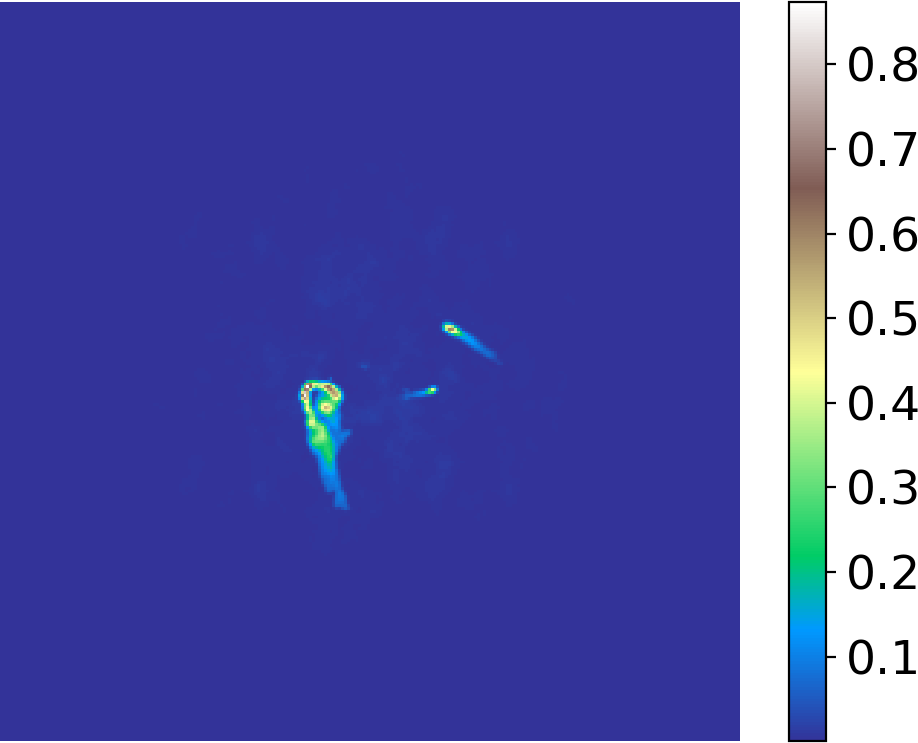} \hfill
   \includegraphics[height=0.35\columnwidth]{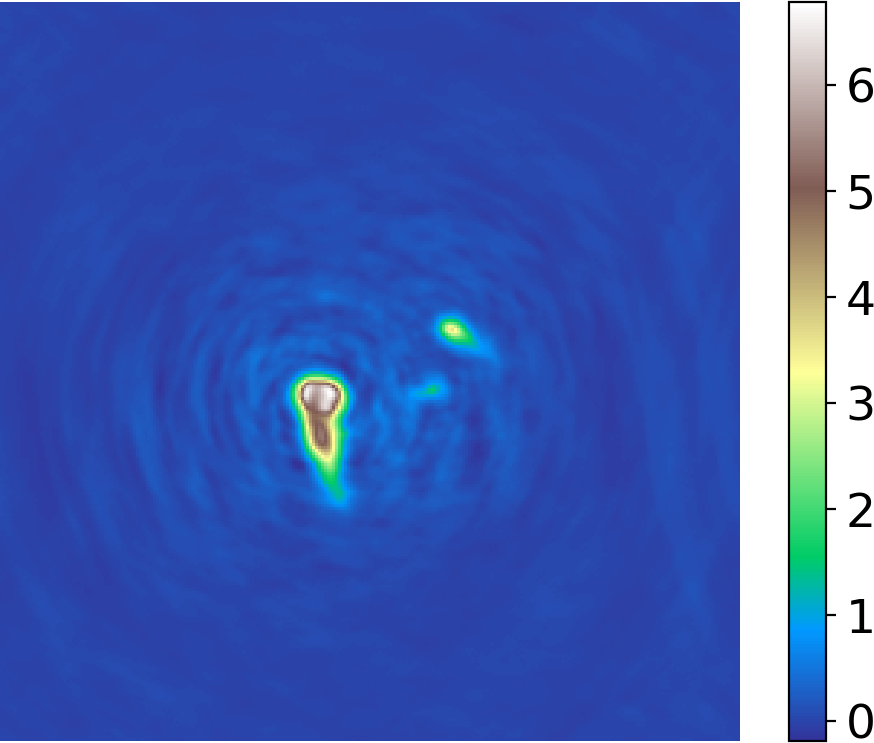} \hfill
    \includegraphics[height=0.35\columnwidth]{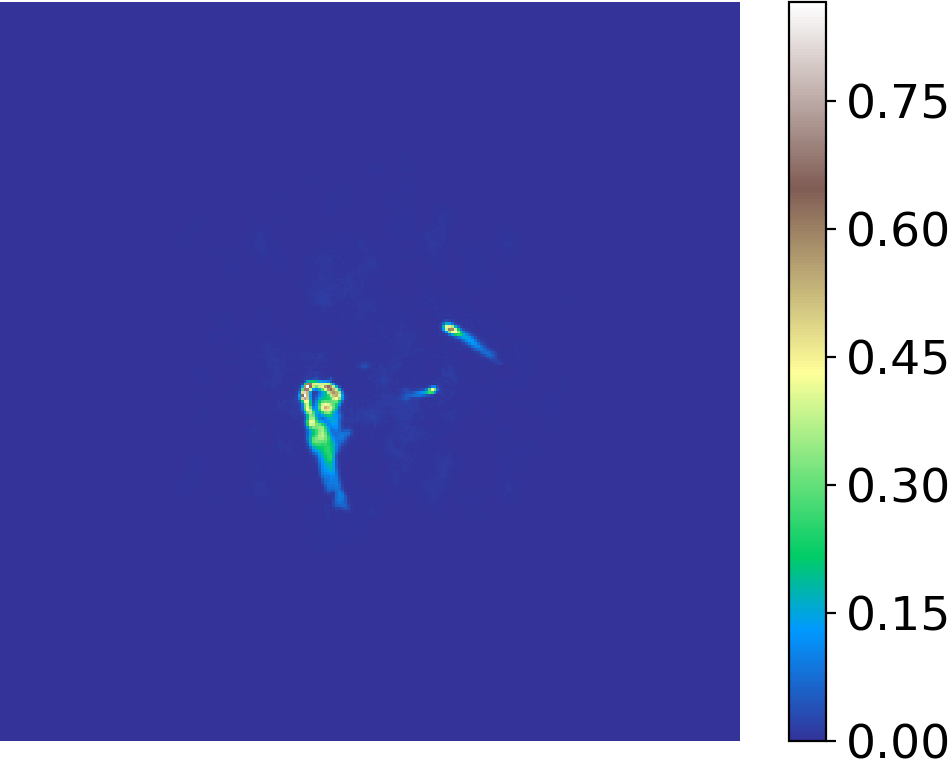} \hfill
   \includegraphics[height=0.35\columnwidth]{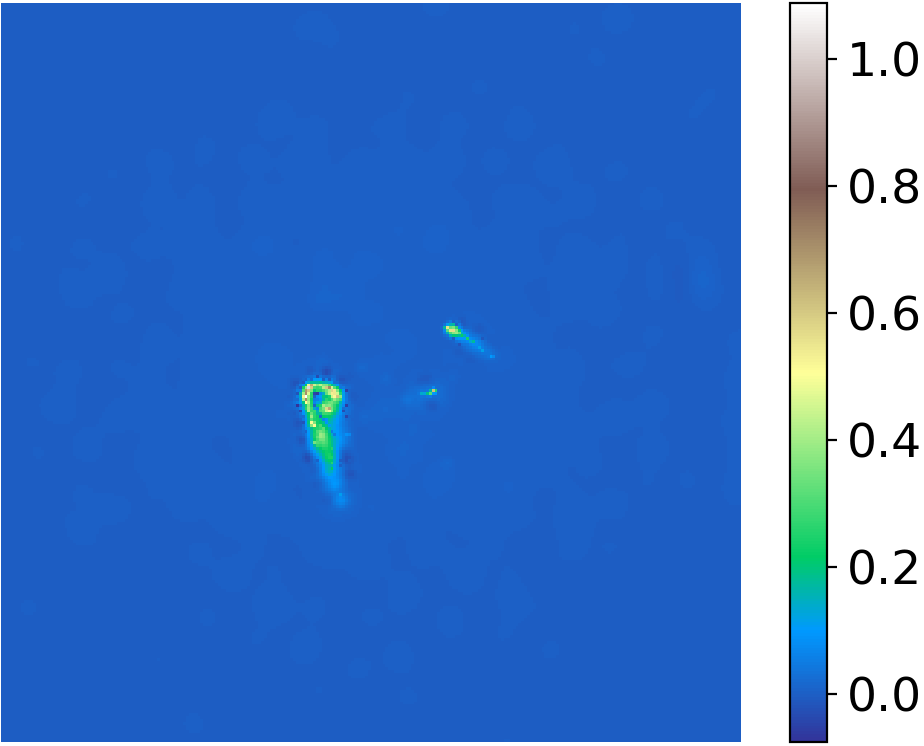}
  }
\medskip 
  \centerline{
   \includegraphics[height=0.35\columnwidth]{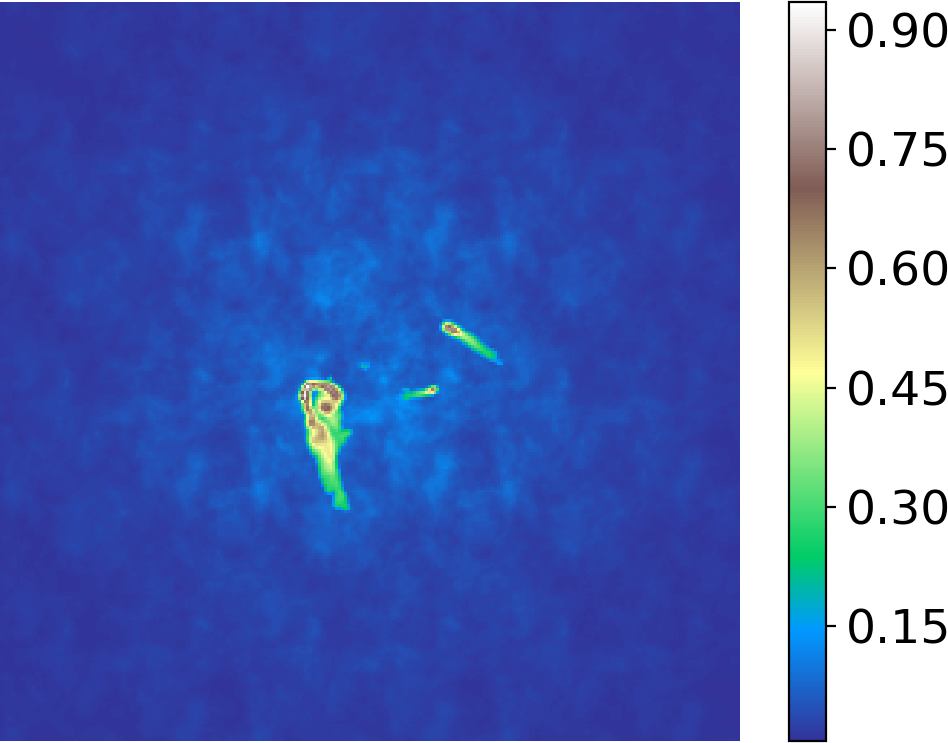} \hfill
   \includegraphics[height=0.35\columnwidth]{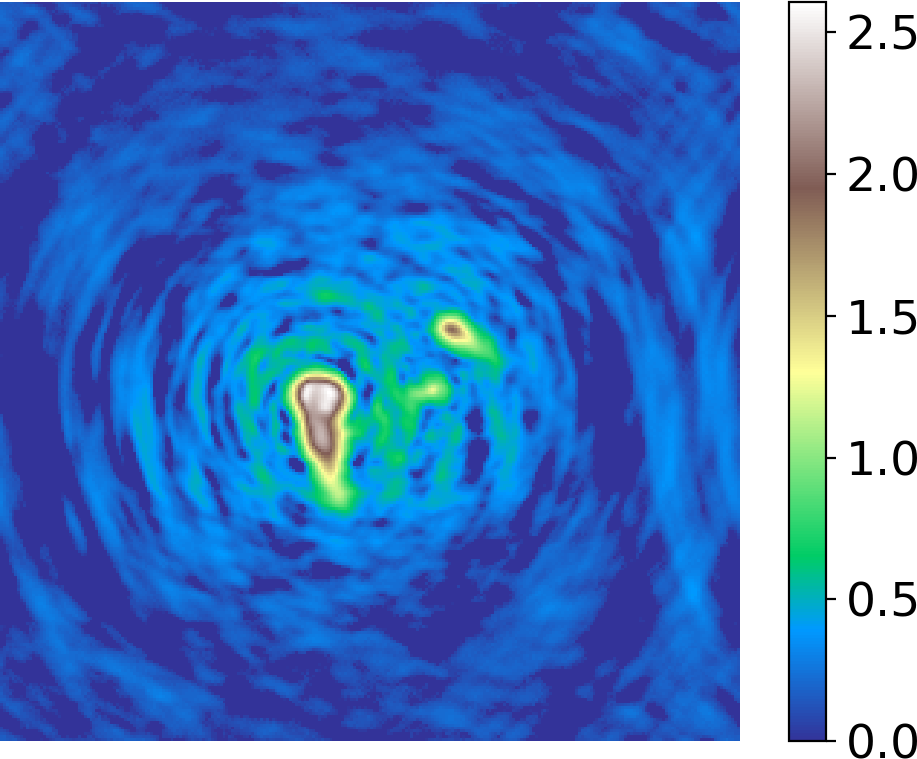} \hfill
   \includegraphics[height=0.35\columnwidth]{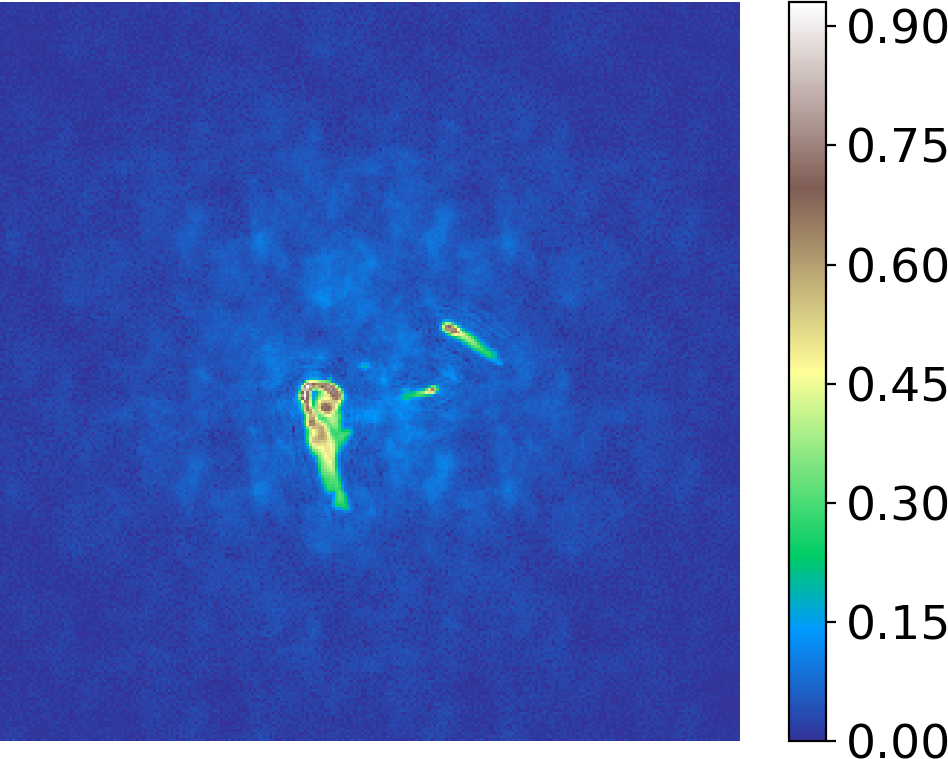} \hfill
    \includegraphics[height=0.35\columnwidth]{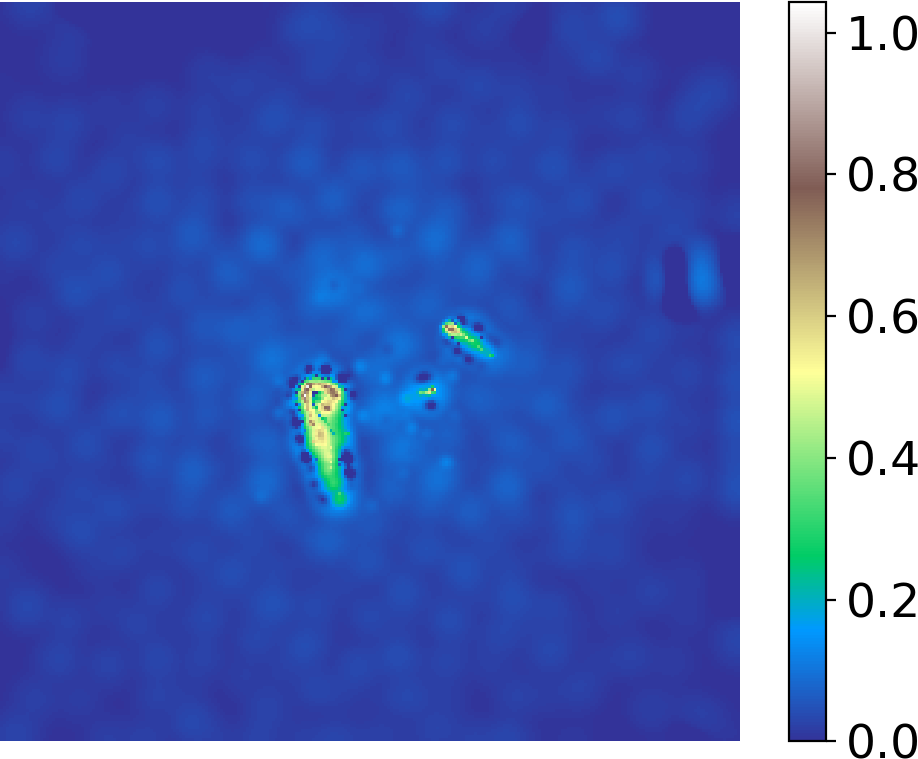}
  }

\caption{First column: Halo sky image. Second column: dirty image. Third column: Halo reconstructed image using MUFFIN. Fourth column: Halo reconstructed image using MS-MFS. All images are shown at the first frequency band. First row shows the images intensities at linear scale while the second row shows the images intensities to the power 0.5 in order to better highlight the small intensities.}
\label{fig:haloresim}
\end{figure*}

Figure \ref{fig:haloresim} compares the performance of the proposed approach with  MS-MFS. The third (MUFFIN) and fourth  (MS-MFS) columns lead to the same conclusion as for the Galaxy-H{\small I} data set. Note that the SNR of the reconstructed image cube is $31.44$ dB for MUFFIN and $14.06$ for MS-MFS.

\begin{figure}
  \centerline{
    \includegraphics[height=0.35\columnwidth]{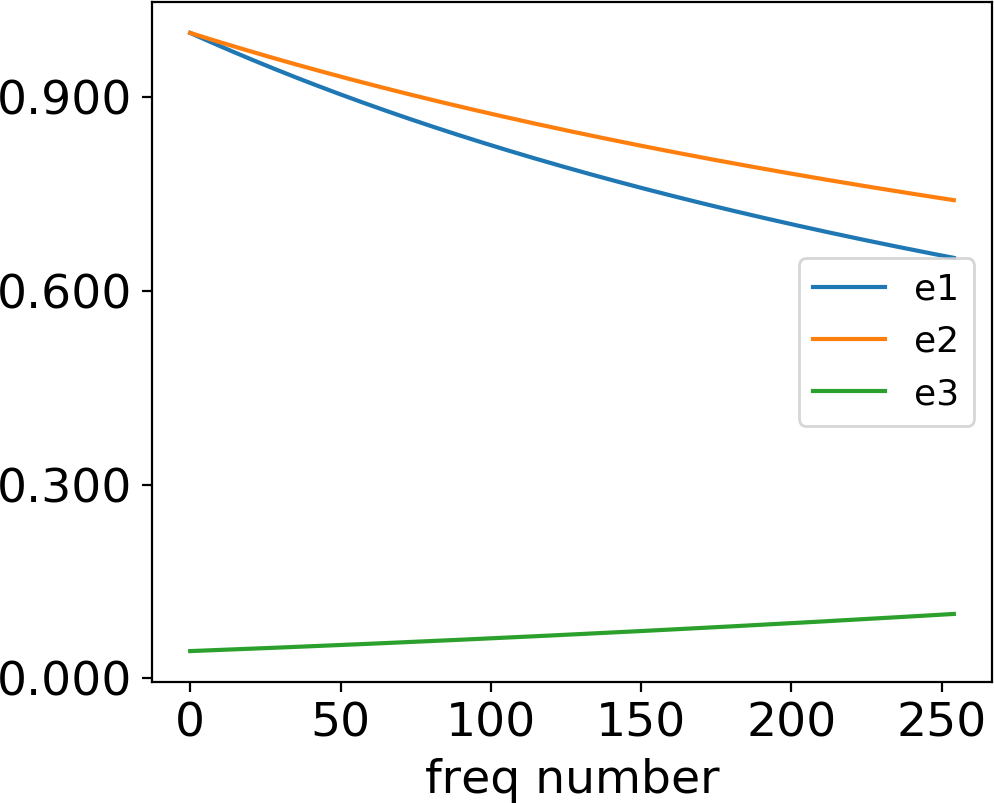}\hfill
    \includegraphics[height=0.35\columnwidth]{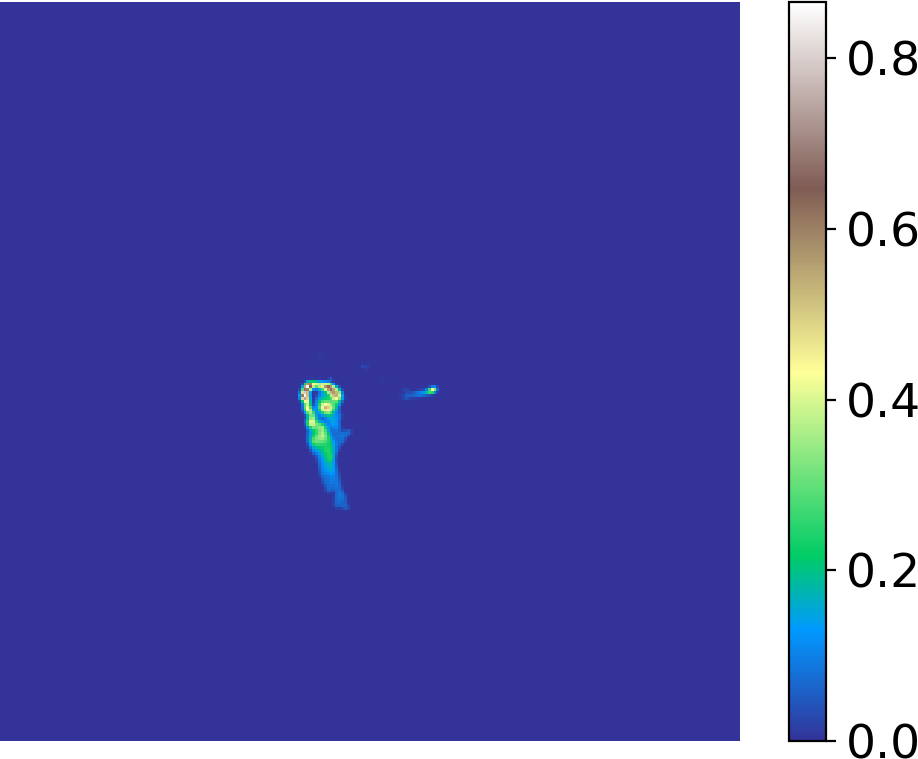}}    
   \medskip   
  \centerline{
    \includegraphics[height=0.35\columnwidth]{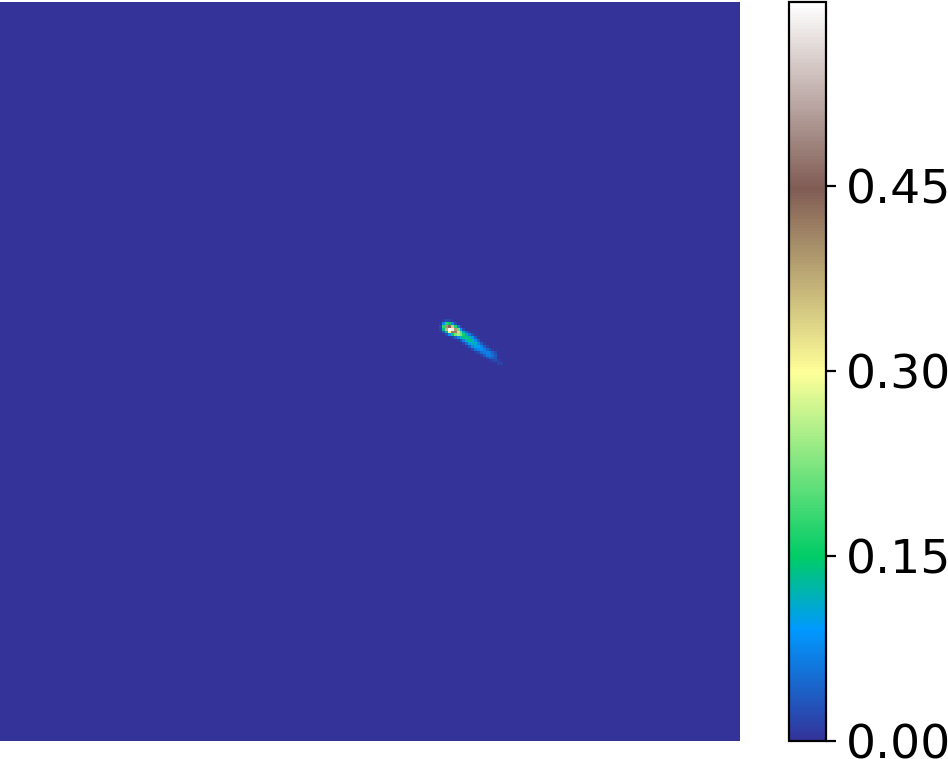}\hfill
    \includegraphics[height=0.35\columnwidth]{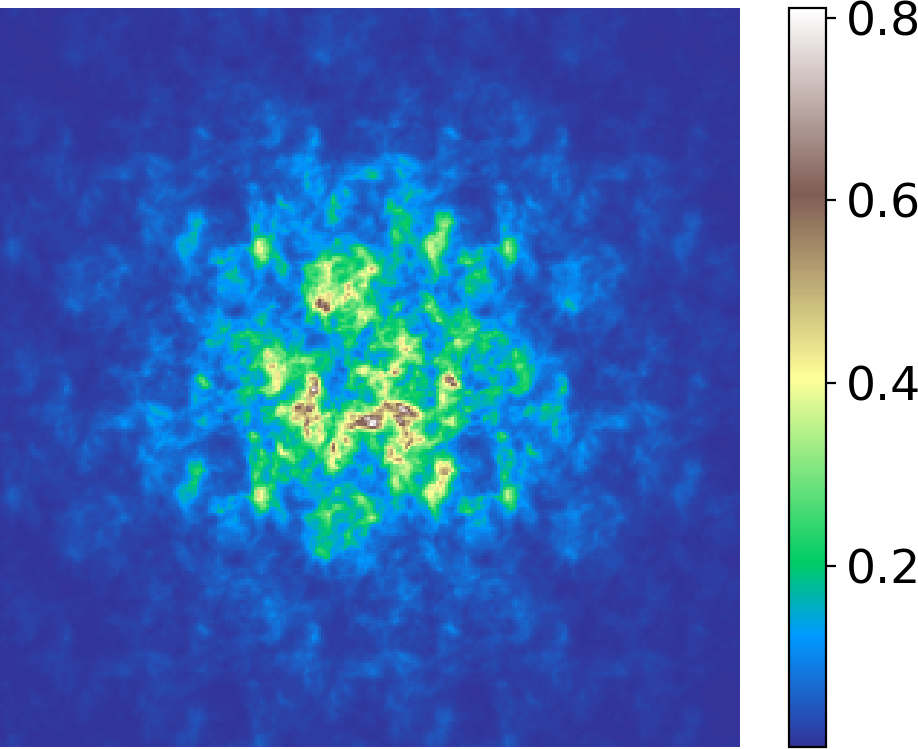}
  }
\caption{`Halo' data set. Top left panel: the endmembers spectra $\be_k$, $k=1,2,3$, corresponding to the three sources. Top right, bottom left and bottom right panels: abundance maps for the two compact sources and the Halo respectively.}
\label{fig:HaloEndAbd}
\end{figure}

\begin{figure*}
  \centerline{
   \includegraphics[height=0.35\columnwidth]{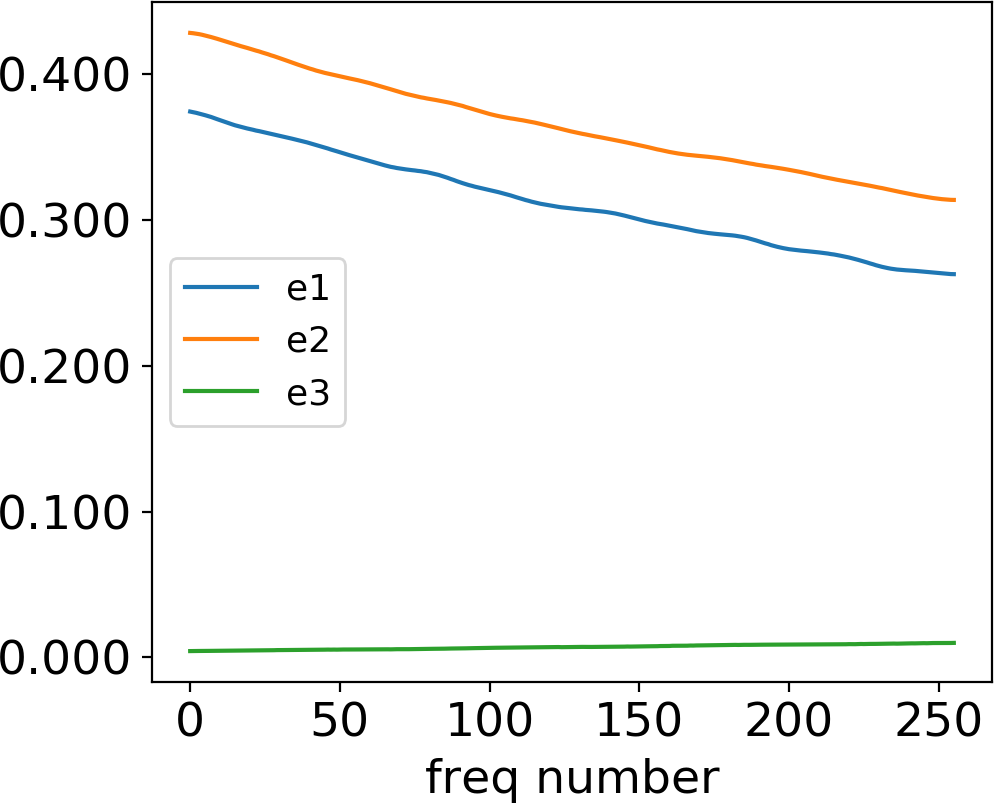} \hfill
   \includegraphics[height=0.35\columnwidth]{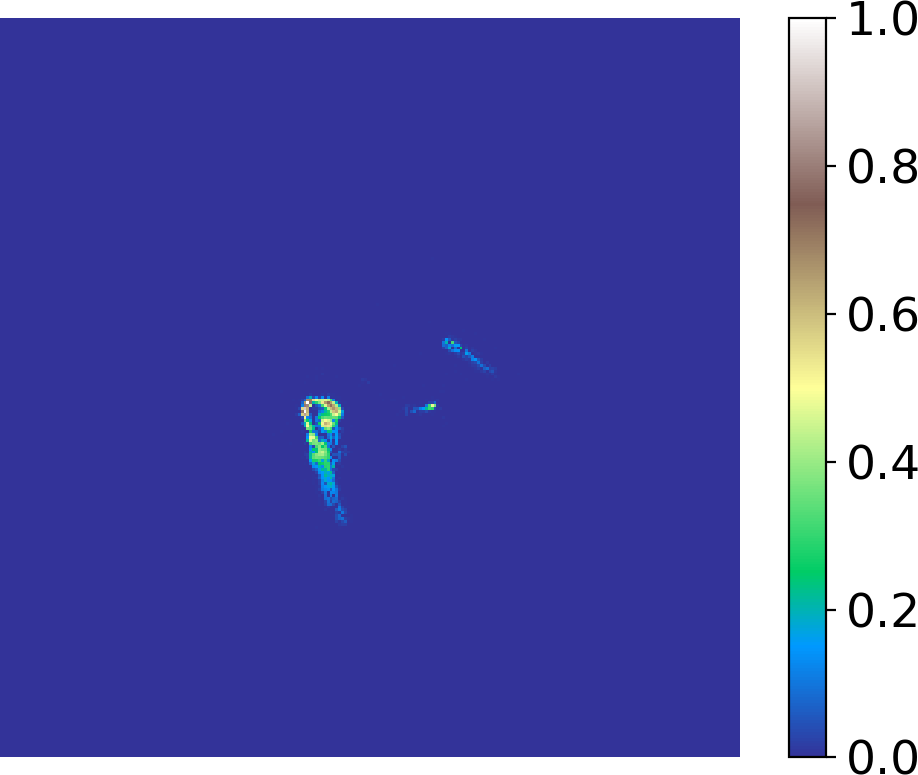} \hfill
    \includegraphics[height=0.35\columnwidth]{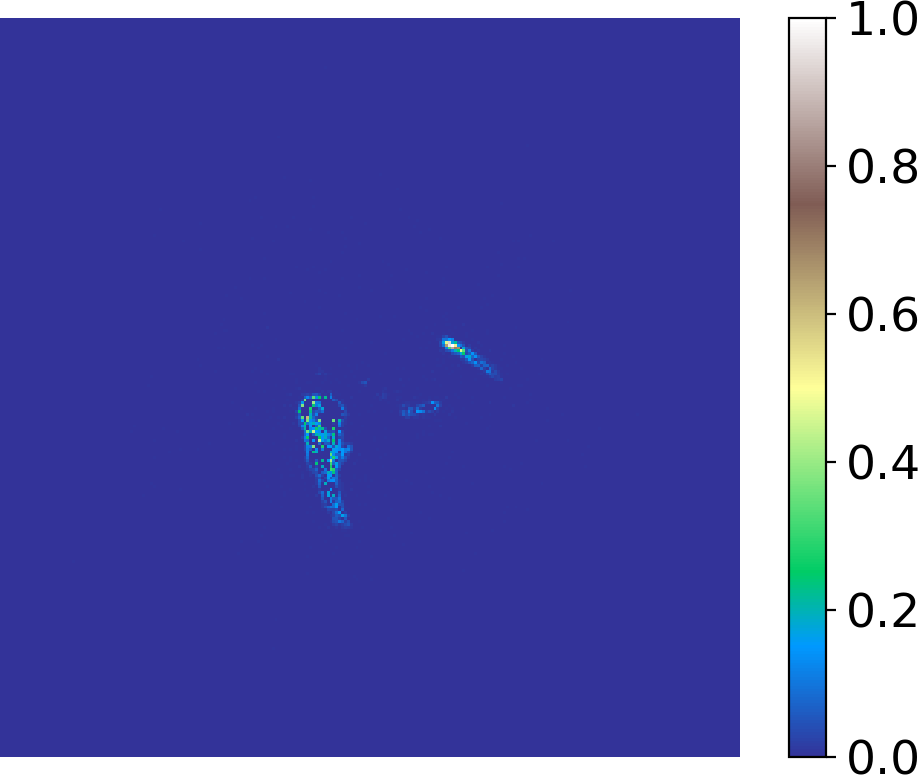} \hfill
    \includegraphics[height=0.35\columnwidth]{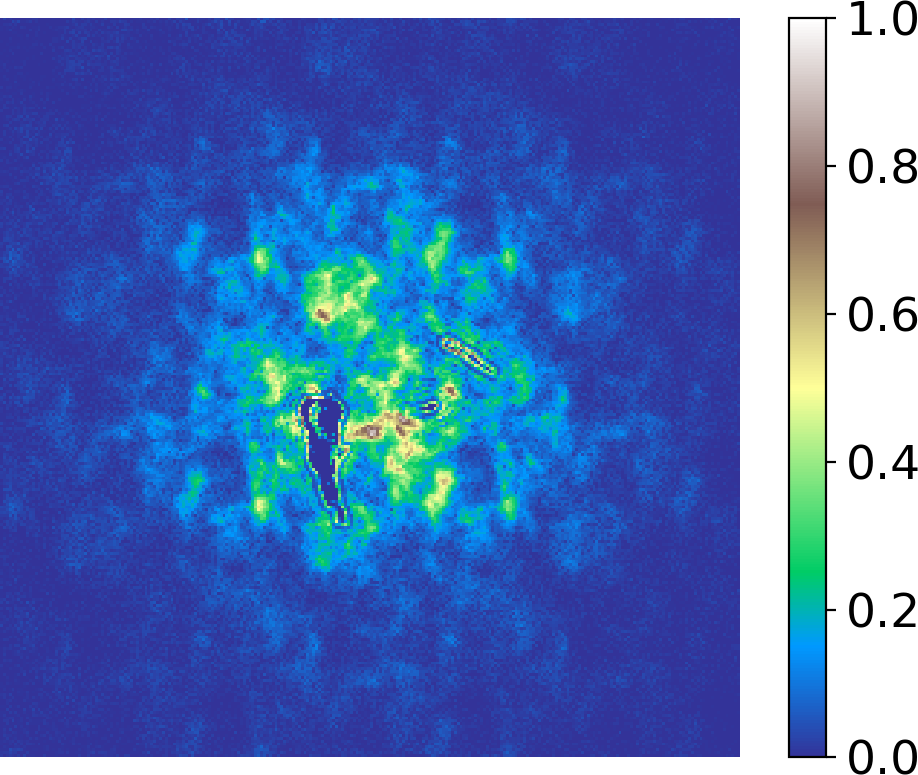} 
  }
  \medskip
  \centerline{
   \includegraphics[height=0.35\columnwidth]{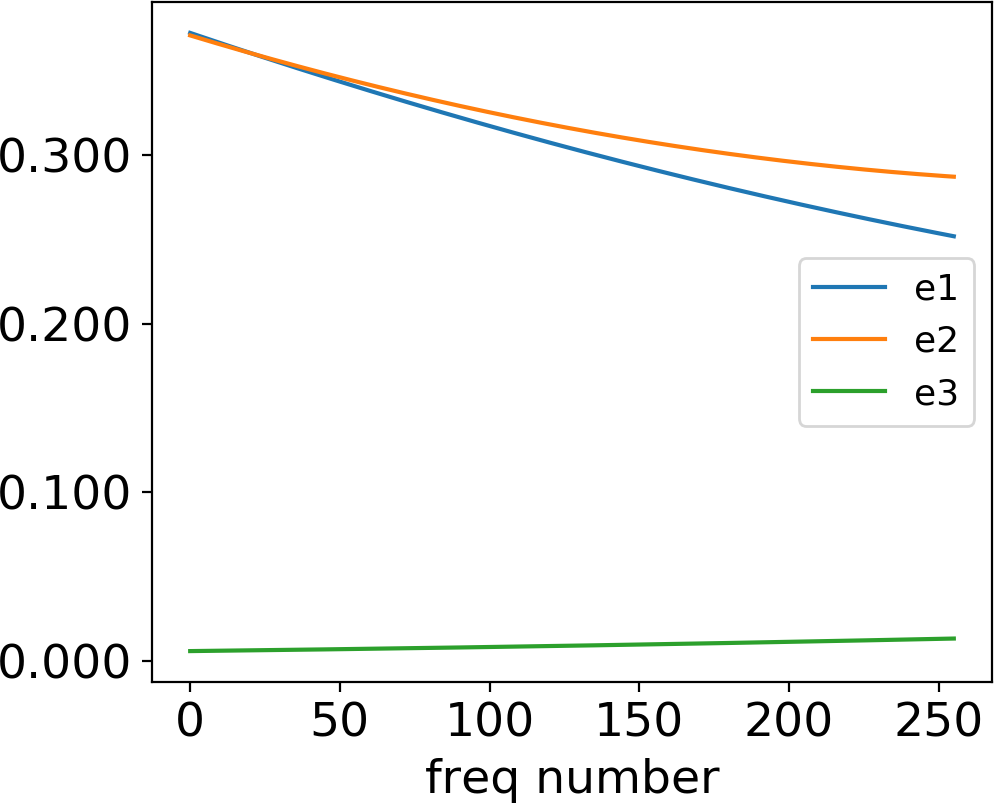} \hfill
   \includegraphics[height=0.35\columnwidth]{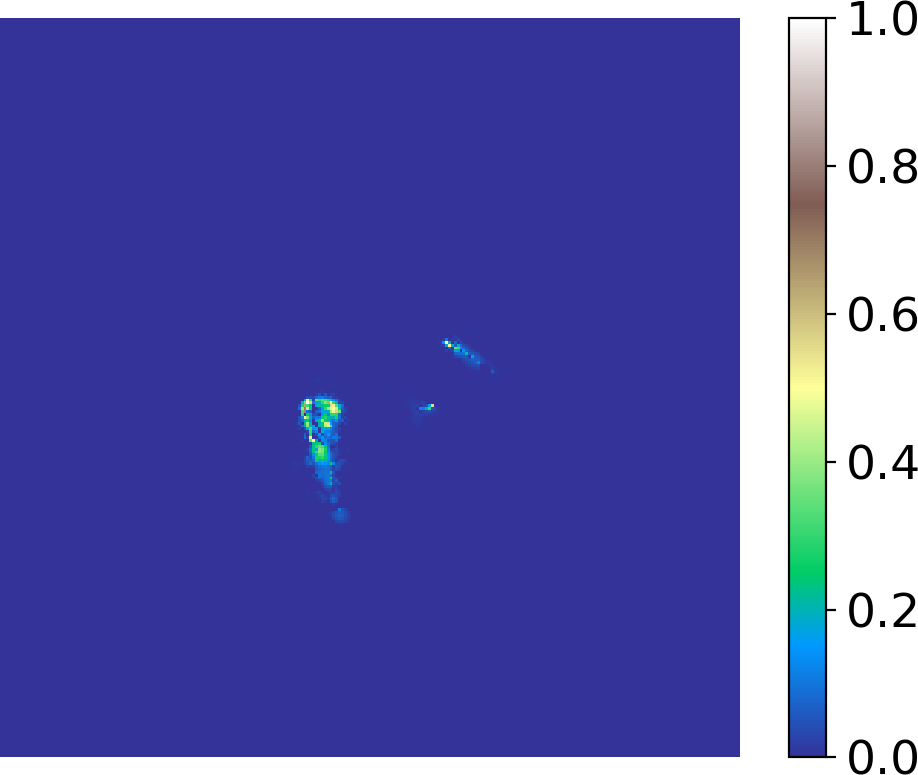} \hfill
    \includegraphics[height=0.35\columnwidth]{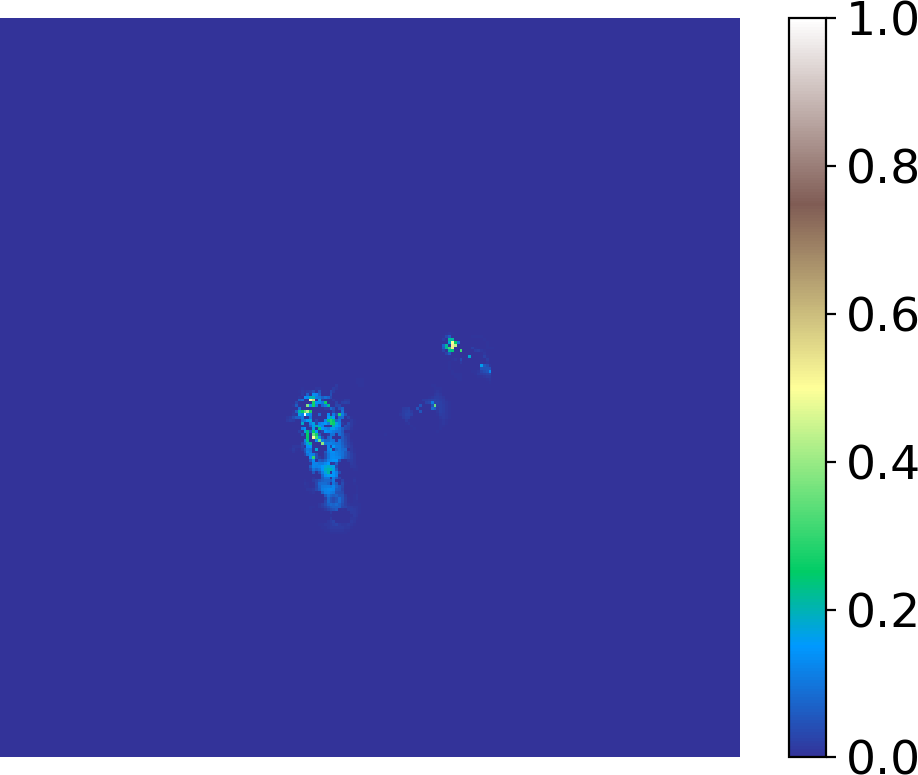} \hfill
    \includegraphics[height=0.35\columnwidth]{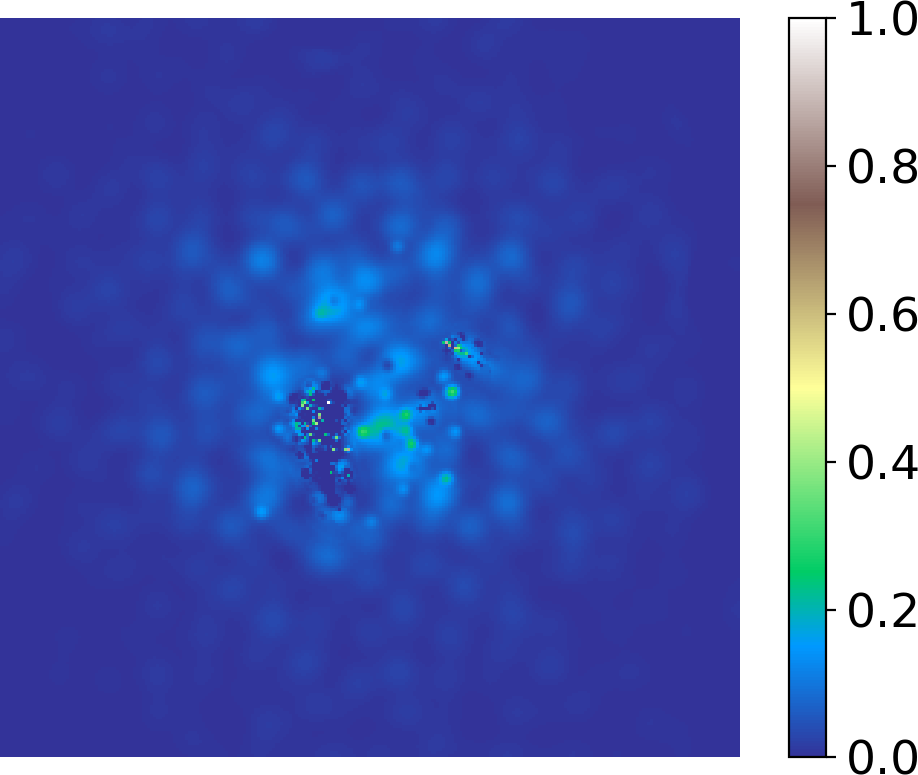} 
  }

\caption{Estimated abundance maps. First column shows the selected spectra from the reconstructed image, the second, third and fourth columns show the abundance maps obtained by solving the positively constrained LS problem. The first and second row correspond to the results obtained with MUFFIN and MS-MFS respectively. }
\label{fig:halounmixres}
\end{figure*}

Finally, Figure \ref{fig:halounmixres} shows the source separation result. We chose three spectra belonging to brightest pixels: one from the faint halo region, one from the  largest galaxy and one from the  smallest galaxy. The spectra of these pixels are shown in the first column of Figure \ref{fig:halounmixres}. We then solved a positively constrained least squares problem in order to find the contribution of each spectrum for each pixel of the reconstructed data cubes. Columns $2$ to $4$ show the recovered abundance maps (to be compared to the maps shown in Fig.\ref{fig:HaloEndAbd}): the first row  corresponds to MUFFIN and the second row  to MS-MFS. It can be seen that in the case of MUFFIN, the two galaxies are better separated than in the MS-MFS case. As for the Halo, it is well separated in both cases, but for MS-MFS  it presents larger artifacts and  differences with respect to the reference. 

%%%%%%%%%%%%%%%%%%% FIGURES %%%%%%%%%%%%%%%%%%%

\section{Conclusions}

In this work, we have developed a spatial-spectral deconvolution method for large multispectral data named MUFFIN. For this algorithm, we designed a specific parallel  implementation and  an original procedure for automatically self-tuning the regularization parameters.  We demonstrated the efficiency of the proposed deconvolution method on simulated but realistic astronomical multispectral images.   The parallel strategy and the SURE-based automatic parameter selection were specifically adapted to the spatial-spectral reconstruction strategy of the MUFFIN algorithm. However, the tools used in this work should provide a foundation for developing analogous parallel implementations and automated parameter selection procedures in the case of different deconvolution problems.

\section*{Acknowledgements}

This work was supported  by the Agence Nationale pour la Recherche, France, (MAGELLAN project, ANR-14-CE23-0004-01). The numerical simulations in this work were performed at the Centre for Intensive Computing `M\'esocentre sigamm' hosted by the Observatoire de la C\^ote dAzur. The authors would like to thank Alain Miniussi for his help on the cluster and P. Serra from Osservatorio Astronomico di Cagliari for his help on the model of the HI line. The authors are very grateful to the referee,  Stephen Ord, for his insightful suggestions and helpful comments to improve the quality of the original manuscript.

\bibliographystyle{mnras}
\bibliography{MyBiblio} % if your bibtex file is called example.bib

\begin{thebibliography}{}
\makeatletter
\relax
\def\mn@urlcharsother{\let\do\@makeother \do\$\do\&\do\#\do\^\do\_\do\%\do\~}
\def\mn@doi{\begingroup\mn@urlcharsother \@ifnextchar [ {\mn@doi@}
  {\mn@doi@[]}}
\def\mn@doi@[#1]#2{\def\@tempa{#1}\ifx\@tempa\@empty \href
  {http://dx.doi.org/#2} {doi:#2}\else \href {http://dx.doi.org/#2} {#1}\fi
  \endgroup}
\def\mn@eprint#1#2{\mn@eprint@#1:#2::\@nil}
\def\mn@eprint@arXiv#1{\href {http://arxiv.org/abs/#1} {{\tt arXiv:#1}}}
\def\mn@eprint@dblp#1{\href {http://dblp.uni-trier.de/rec/bibtex/#1.xml}
  {dblp:#1}}
\def\mn@eprint@#1:#2:#3:#4\@nil{\def\@tempa {#1}\def\@tempb {#2}\def\@tempc
  {#3}\ifx \@tempc \@empty \let \@tempc \@tempb \let \@tempb \@tempa \fi \ifx
  \@tempb \@empty \def\@tempb {arXiv}\fi \@ifundefined
  {mn@eprint@\@tempb}{\@tempb:\@tempc}{\expandafter \expandafter \csname
  mn@eprint@\@tempb\endcsname \expandafter{\@tempc}}}

\bibitem[\protect\citeauthoryear{Abdulaziz, Dabbech, Onose  \& Wiaux}{Abdulaziz
  et~al.}{2016}]{abdulaziz2016low}
Abdulaziz A.,  Dabbech A.,  Onose A.,   Wiaux Y.,  2016, in Signal Processing
  Conference (EUSIPCO), 2016 24th European. pp 388--392

\bibitem[\protect\citeauthoryear{Ammanouil, Ferrari, Flamary, Ferrari  \&
  Mary}{Ammanouil et~al.}{2017}]{ammanouil2017multi}
Ammanouil R.,  Ferrari A.,  Flamary R.,  Ferrari C.,   Mary D.,  2017, European
  Signal Processing Conference (EUSIPCO)

\bibitem[\protect\citeauthoryear{Ammanouil, Ferrari  \& Richard}{Ammanouil
  et~al.}{2018}]{ammanouilEusipco2017}
Ammanouil R.,  Ferrari A.,   Richard C.,  2018, in IEEE International
  Conference on Acoustics, Speech and Signal Processing.

\bibitem[\protect\citeauthoryear{Avron \& Toledo}{Avron \&
  Toledo}{2011}]{avron2011randomized}
Avron H.,  Toledo S.,  2011, Journal of the ACM (JACM), 58, 8

\bibitem[\protect\citeauthoryear{Bajkova \& Pushkarev}{Bajkova \&
  Pushkarev}{2011}]{bajkova2011multifrequency}
Bajkova A.~T.,  Pushkarev A.~B.,  2011, \mnras, 417, 434

\bibitem[\protect\citeauthoryear{Beck \& Teboulle}{Beck \&
  Teboulle}{2009}]{Beck2009}
Beck A.,  Teboulle M. I. P. I. T.~o.,  2009, Image Processing, IEEE
  Transactions on, 18

\bibitem[\protect\citeauthoryear{Bhatnagar, Cornwell, Golap  \& Uson}{Bhatnagar
  et~al.}{2008}]{bhatnagar2008}
Bhatnagar S.,  Cornwell T.~J.,  Golap K.,   Uson J.~M.,  2008, \aap, 487, 419

\bibitem[\protect\citeauthoryear{Boyd, Parikh, Chu, Peleato  \& Eckstein}{Boyd
  et~al.}{2011}]{boyd2011distributed}
Boyd S.,  Parikh N.,  Chu E.,  Peleato B.,   Eckstein J.,  2011, Foundations
  and Trends in Machine learning, 3, 1

\bibitem[\protect\citeauthoryear{Cand{\`e}s \& Wakin}{Cand{\`e}s \&
  Wakin}{2008}]{candes2008introduction}
Cand{\`e}s E.,  Wakin M.,  2008, IEEE signal processing magazine, 25, 21

\bibitem[\protect\citeauthoryear{Cand{\`{e}}s, Wakin  \& Boyd}{Cand{\`{e}}s
  et~al.}{2008}]{Cands2008}
Cand{\`{e}}s E.~J.,  Wakin M.~B.,   Boyd S.~P.,  2008, Journal of Fourier
  Analysis and Applications, 14, 877

\bibitem[\protect\citeauthoryear{Carrillo, McEwen  \& Wiaux}{Carrillo
  et~al.}{2014}]{Purify}
Carrillo R.~E.,  McEwen J.~D.,   Wiaux Y.,  2014, \mnras, 439, 3591

\bibitem[\protect\citeauthoryear{Chambolle \& Pock}{Chambolle \&
  Pock}{2010}]{Chambolle2010}
Chambolle A.,  Pock T.,  2010, Journal of Mathematical Imaging and Vision, 40,
  120

\bibitem[\protect\citeauthoryear{Combettes \& Pesquet}{Combettes \&
  Pesquet}{2011}]{combettes2011}
Combettes P.~L.,  Pesquet J.-C.,  2011, in , Fixed-Point Algorithms for Inverse
  Problems in Science and Engineering.
Springer New York, New York, NY, pp 185--212

\bibitem[\protect\citeauthoryear{Combettes \& Pesquet}{Combettes \&
  Pesquet}{2012}]{combettes2012primal}
Combettes P.,  Pesquet J.-C.~P.,  2012, Set-Valued and variational analysis,
  20, 307

\bibitem[\protect\citeauthoryear{Condat}{Condat}{2014}]{Condat}
Condat L.,  2014, IEEE Signal Processing Letters, 21, 985

\bibitem[\protect\citeauthoryear{Conway, Cornwell  \& Wilkinson}{Conway
  et~al.}{1990}]{Conway90}
Conway J.~E.,  Cornwell T.~J.,   Wilkinson P.~N.,  1990, \mnras, 246, 490

\bibitem[\protect\citeauthoryear{Cornwell}{Cornwell}{2008}]{Cornwell2008}
Cornwell T.~J.,  2008, {IEEE} Journal of Selected Topics in Signal Processing,
  2, 793

\bibitem[\protect\citeauthoryear{Dabbech, Ferrari, Mary, Slezak, Smirnov  \&
  Kenyon}{Dabbech et~al.}{2015}]{dabbech15}
Dabbech A.,  Ferrari C.,  Mary D.,  Slezak E.,  Smirnov O.,   Kenyon J.~S.,
  2015, \aap, 576, A7

\bibitem[\protect\citeauthoryear{Dabbech, Onose, Abdulaziz, Perley, Smirnov  \&
  Wiaux}{Dabbech et~al.}{2018}]{dabbech2018}
Dabbech A.,  Onose A.,  Abdulaziz A.,  Perley R.~A.,  Smirnov O.~M.,   Wiaux
  Y.,  2018, \mnras, 476, 2853

\bibitem[\protect\citeauthoryear{Deledalle, Vaiter, Fadili  \&
  Peyr{\'e}}{Deledalle et~al.}{2014}]{deledalle2014stein}
Deledalle C.-A.,  Vaiter S.,  Fadili J.,   Peyr{\'e} G.,  2014, SIAM Journal on
  Imaging Sciences, 7, 2448

\bibitem[\protect\citeauthoryear{Dewdney, Hall, Schilizzi, Lazio  \&
  Joseph}{Dewdney et~al.}{2009}]{dewdney2009square}
Dewdney P.,  Hall P.,  Schilizzi R.,  Lazio T.,   Joseph L.~W.,  2009,
  Proceedings of the IEEE, 97, 1482

\bibitem[\protect\citeauthoryear{Dewdney, Turner, Millenaar, McCool, Lazio  \&
  Cornwell}{Dewdney et~al.}{2013}]{dewdney2013ska1}
Dewdney P.,  Turner W.,  Millenaar R.,  McCool R.,  Lazio J.,   Cornwell T.,
  2013, Technical Report~1, SKA1 system baseline design.
SKA Office

\bibitem[\protect\citeauthoryear{Donoho}{Donoho}{2006}]{donoho2006compressed}
Donoho D.~L.,  2006, IEEE Transactions on information theory, 52, 1289

\bibitem[\protect\citeauthoryear{Elad, Milanfar  \& Rubinstein}{Elad
  et~al.}{2007}]{elad2007analysis}
Elad M.,  Milanfar P.,   Rubinstein R.,  2007, Inverse problems, 23, 947

\bibitem[\protect\citeauthoryear{Eldar}{Eldar}{2009}]{eldar2009generalized}
Eldar Y.~C.,  2009, IEEE Transactions on Signal Processing, 57, 471

\bibitem[\protect\citeauthoryear{{Feretti}, {Giovannini}, {Govoni}  \&
  {Murgia}}{{Feretti} et~al.}{2012}]{Feretti12}
{Feretti} L.,  {Giovannini} G.,  {Govoni} F.,   {Murgia} M.,  2012, \aapr, 20,
  54

\bibitem[\protect\citeauthoryear{{Ferrari}, {Govoni}, {Schindler}, {Bykov}  \&
  {Rephaeli}}{{Ferrari} et~al.}{2008}]{CFerrari2008}
{Ferrari} C.,  {Govoni} F.,  {Schindler} S.,  {Bykov} A.~M.,   {Rephaeli} Y.,
  2008, Space Sci. Rev., 134, 93

\bibitem[\protect\citeauthoryear{{Ferrari} et~al.,}{{Ferrari}
  et~al.}{2015}]{CFerrari2015}
{Ferrari} C.,  et~al., 2015, Advancing Astrophysics with the Square Kilometre
  Array (AASKA14), p.~75

\bibitem[\protect\citeauthoryear{Garsden, JN.~Girard  et~al.}{Garsden
  et~al.}{2015}]{garsden15}
Garsden H.,  JN.~Girard J.,   et~al., 2015, \aap, 575, A90

\bibitem[\protect\citeauthoryear{Giryes, Elad  \& Eldar}{Giryes
  et~al.}{2011}]{giryes2011projected}
Giryes R.,  Elad M.,   Eldar Y.~C.,  2011, Applied and Computational Harmonic
  Analysis, 30, 407

\bibitem[\protect\citeauthoryear{Golub, Heath  \& Wahba}{Golub
  et~al.}{1979}]{golub1979generalized}
Golub G.~H.,  Heath M.,   Wahba G.,  1979, Technometrics, 21, 215

\bibitem[\protect\citeauthoryear{Hansen \& O'Leary}{Hansen \&
  O'Leary}{1993}]{hansen1993use}
Hansen P.~C.,  O'Leary D.~P.,  1993, SIAM Journal on Scientific Computing, 14,
  1487

\bibitem[\protect\citeauthoryear{H{\"o}gbom}{H{\"o}gbom}{1974}]{hogbom1974aperture}
H{\"o}gbom J.~A.,  1974, \aap Supplement Series, 15, 417

\bibitem[\protect\citeauthoryear{Jiang, Bobin  \& Starck}{Jiang
  et~al.}{2017}]{jiang2017joint}
Jiang M.,  Bobin J.,   Starck J.-L.,  2017, SIAM Journal on Imaging Sciences,
  10, 1997

\bibitem[\protect\citeauthoryear{Jongerius, Wijnholds, Nijboer  \&
  Corporaal}{Jongerius et~al.}{2014}]{jongerius2014}
Jongerius R.,  Wijnholds S.,  Nijboer R.,   Corporaal H.,  2014, Computer, 47,
  48

\bibitem[\protect\citeauthoryear{Junklewitz, Bell  \& En{\ss}lin}{Junklewitz
  et~al.}{2015}]{Junklewitz2015}
Junklewitz H.,  Bell M.,   En{\ss}lin T.,  2015, \aap, 581, A59

\bibitem[\protect\citeauthoryear{Karl}{Karl}{2005}]{karl2000regularization}
Karl W.~C.,  2005, Handbook of image video processing, pp 183--202

\bibitem[\protect\citeauthoryear{Kartik, Carrillo, Thiran  \& Wiaux}{Kartik
  et~al.}{2017}]{vijay2017fourier}
Kartik S.~V.,  Carrillo R.,  Thiran J.-P.,   Wiaux Y.,  2017, \mnras, 468, 2382

\bibitem[\protect\citeauthoryear{Kellermann}{Kellermann}{1974}]{Kellerman74}
Kellermann K.,  1974, in Verschuur G.,  Kellermann K.,  eds, , Galactic and
  Extragalactic Radio Astronomy.
Springer-Verlag, Chapt. Radio galaxies and quasars

\bibitem[\protect\citeauthoryear{Kogan \& Greisen}{Kogan \&
  Greisen}{2009}]{kogan2009}
Kogan L.,  Greisen E.~W.,  2009, Technical report, {Faceted imaging in AIPS}.
{AIPS Memos}

\bibitem[\protect\citeauthoryear{Kraus}{Kraus}{1986}]{krausb}
Kraus J.~D.,  1986, Radio {A}stronomy.
Cygnus-Quasar, Powell, Ohio

\bibitem[\protect\citeauthoryear{Lanusse, Starck, Woiselle  \& Fadili}{Lanusse
  et~al.}{2014}]{Lanusse2014}
Lanusse F.,  Starck J.-L.,  Woiselle A.,   Fadili M.,  2014, in , Advances in
  Imaging and Electron Physics.
Elsevier, pp 99--204

\bibitem[\protect\citeauthoryear{Loi, Murgia, Govoni, Vacca, Prandoni, Li,
  Feretti  \& Giovannini}{Loi et~al.}{2018}]{loi18}
Loi F.,  Murgia M.,  Govoni F.,  Vacca V.,  Prandoni I.,  Li H.,  Feretti L.,
  Giovannini G.,  2018, Galaxies, 6, 133

\bibitem[\protect\citeauthoryear{McEwen \& Wiaux}{McEwen \&
  Wiaux}{2011}]{mcewen2011compressed}
McEwen J.,  Wiaux Y.,  2011, in Image Processing (ICIP), 2011 18th IEEE
  International Conference on. pp 1313--1316

\bibitem[\protect\citeauthoryear{Meillier, Ammanouil, Ferrari  \&
  Bianchi}{Meillier et~al.}{2018}]{meillier2018distribution}
Meillier C.,  Ammanouil R.,  Ferrari A.,   Bianchi P.,  2018, Signal
  Processing: Image Communication, 67, 149

\bibitem[\protect\citeauthoryear{Morozov}{Morozov}{1966}]{morozov1966solution}
Morozov V.~A.,  1966, in Soviet Math. Dokl. pp 414--417

\bibitem[\protect\citeauthoryear{Murgia, Govoni, Feretti, Giovannini,
  Dallacasa, Fanti, Taylor  \& Dolag}{Murgia et~al.}{2004}]{murgia04}
Murgia M.,  Govoni F.,  Feretti L.,  Giovannini G.,  Dallacasa D.,  Fanti R.,
  Taylor G.~B.,   Dolag K.,  2004, \aap, 424, 429

\bibitem[\protect\citeauthoryear{Noordam \& Smirnov}{Noordam \&
  Smirnov}{2010}]{noordam2010meqtrees}
Noordam J.~E.,  Smirnov O.~M.,  2010, \aap, 524, A61

\bibitem[\protect\citeauthoryear{Onose, Carrillo, McEwen  \& Wiaux}{Onose
  et~al.}{2016a}]{onose2016randomised}
Onose A.,  Carrillo R.~E.,  McEwen J.~D.,   Wiaux Y.,  2016a, in Signal
  Processing Conference (EUSIPCO), 2016 24th European. pp 1448--1452

\bibitem[\protect\citeauthoryear{Onose, Carrillo, Repetti  et~al.}{Onose
  et~al.}{2016b}]{Onose16}
Onose A.,  Carrillo R.~E.,  Repetti A.,   et~al., 2016b, \mnras, 462, 4314

\bibitem[\protect\citeauthoryear{Paris, Suleiman, Mary  \& Ferrari}{Paris
  et~al.}{2013}]{Paris2013}
Paris S.,  Suleiman R.,  Mary D.,   Ferrari A.,  2013, in 2013 {IEEE}
  International Conference on Acoustics, Speech and Signal Processing. {IEEE}

\bibitem[\protect\citeauthoryear{Ramani, Blu  \& Unser}{Ramani
  et~al.}{2008}]{ramani2008blind}
Ramani S.,  Blu T.,   Unser M.,  2008, in IEEE International Conference on
  Acoustics, Speech and Signal Processing. pp 905--908

\bibitem[\protect\citeauthoryear{Ramani, Liu, Rosen, Nielsen  \&
  Fessler}{Ramani et~al.}{2012}]{ramani2012regularization}
Ramani S.,  Liu Z.,  Rosen J.,  Nielsen J.-F.,   Fessler J.~A.,  2012, IEEE
  Transactions on Image Processing, 21, 3659

\bibitem[\protect\citeauthoryear{Rau \& Cornwell}{Rau \&
  Cornwell}{2011}]{rau2011multi}
Rau U.,  Cornwell T.~J.,  2011, \aap, 532, A71

\bibitem[\protect\citeauthoryear{Rau, Bhatnagar, Voronkov  \& Cornwell}{Rau
  et~al.}{2009}]{rau2009}
Rau U.,  Bhatnagar S.,  Voronkov M.~A.,   Cornwell T.~J.,  2009, Proceedings of
  the IEEE, 97, 1472

\bibitem[\protect\citeauthoryear{Regi{\'n}ska}{Regi{\'n}ska}{1996}]{reginska1996regularization}
Regi{\'n}ska T.,  1996, SIAM Journal on Scientific Computing, 17, 740

\bibitem[\protect\citeauthoryear{Renard, Thi{\'e}baut  \& Malbet}{Renard
  et~al.}{2011}]{Renard2011}
Renard S.,  Thi{\'e}baut {\'E}.,   Malbet F.,  2011, \aap, 533, A64

\bibitem[\protect\citeauthoryear{Scaife \& Heald}{Scaife \&
  Heald}{2012}]{Scaife12}
Scaife A.,  Heald G.,  2012, \mnras, 423, L30

\bibitem[\protect\citeauthoryear{Setzer, Steidl  \& Teuber}{Setzer
  et~al.}{2010}]{setzer2010deblurring}
Setzer S.,  Steidl G.,   Teuber T.,  2010, Journal of Visual Communication and
  Image Representation, 21, 193

\bibitem[\protect\citeauthoryear{Starck \& Bobin}{Starck \&
  Bobin}{2010}]{starck2010astronomical}
Starck J.-L.,  Bobin J.,  2010, Proceedings of the IEEE, 98, 1021

\bibitem[\protect\citeauthoryear{Stein}{Stein}{1981}]{stein1981estimation}
Stein C.~M.,  1981, The annals of Statistics, pp 1135--1151

\bibitem[\protect\citeauthoryear{Tasse, van~der Tol, van Zwieten, van Diepen
  \& Bhatnagar}{Tasse et~al.}{2013}]{tasse2013applying}
Tasse C.,  van~der Tol S.,  van Zwieten J.,  van Diepen G.,   Bhatnagar S.,
  2013, \aap, 553, A105

\bibitem[\protect\citeauthoryear{Tasse et~al.,}{Tasse et~al.}{2018}]{tasse2018}
Tasse C.,  et~al., 2018, \aap, 611, A87

\bibitem[\protect\citeauthoryear{V{\~{u}}}{V{\~{u}}}{2011}]{Vu11}
V{\~{u}} B.~C.,  2011, Advances in Computational Mathematics, 38, 667

\bibitem[\protect\citeauthoryear{Wenger \& Magnor}{Wenger \&
  Magnor}{2014}]{Wenger2014}
Wenger S.,  Magnor M.,  2014, Technical report, A Sparse Reconstruction
  Algorithm for Multi-Frequency Radio Images.
Computer Graphics Lab, TU Braunschweig

\bibitem[\protect\citeauthoryear{Wiaux, Jacques, Puy, Scaife  \&
  Vandergheynst}{Wiaux et~al.}{2009}]{wiaux2009compressed}
Wiaux Y.,  Jacques L.,  Puy G.,  Scaife A.,   Vandergheynst P.,  2009, \mnras,
  395, 1733

\bibitem[\protect\citeauthoryear{Wijnholds, van~der Tol, Nijboer  \& van~der
  Veen}{Wijnholds et~al.}{2010}]{wijnholds2010challenges}
Wijnholds S.,  van~der Tol S.,  Nijboer R.,   van~der Veen A.-J.,  2010, Signal
  Processing Magazine, IEEE, 27, 30

\bibitem[\protect\citeauthoryear{van Weeren et~al.,}{van Weeren
  et~al.}{2016}]{vanWeeren2016}
van Weeren R.~J.,  et~al., 2016, \apjs, 223, 1

\makeatother
\end{thebibliography}

% Don't change these lines
\bsp	% typesetting comment
\label{lastpage}
\end{document}